\documentclass[review,3p,times, 10pt]{elsarticle}
\usepackage{multirow}
\usepackage{longtable}
\usepackage{tikz}
\usepackage{pgf-pie}

\usepackage{pgfplots}
\pgfplotsset{width=20cm,compat=1.8}

\usepackage{graphicx}
\usepackage{caption}
\usepackage{subcaption}
\usepackage{url}
\usepackage{comment}

\usepackage{appendix}
\DeclareCaptionLabelFormat{AppendixTables}{A.#2}

\usepackage{amssymb}

\journal{arXiv}

\begin{document}

\begin{frontmatter}



\title{
A Systematic Review of Mobile Apps for Child Sexual Abuse Education: Limitations and Design Guidelines
}


\author[inst1]{Sadia Tasnuva Pritha}
\ead{u1604122@student.cuet.ac.bd}
\author[inst1]{Rahnuma Tasnim}
\ead{u1604103@student.cuet.ac.bd}
\author[inst4]{Muhammad Ashad Kabir\corref{corau}}
\cortext[corau]{Corresponding author}
\ead{akabir@csu.edu.au}
\author[inst2]{Sumaiya Amin}
\ead{x2020gae@stfx.ca}
\author[inst2]{Anik Das}
\ead{x2021gmg@stfx.ca}

\affiliation[inst1]{organization={Department of Computer Science and Engineering, Chittagong University of Engineering and Technology},
            city={Chattogram},
            postcode={4349}, 
            country={Bangladesh}}

\affiliation[inst4]{organization={School of Computing and Mathematics, Charles Sturt University},
            addressline={Panorama Ave}, 
            city={Bathurst},
            postcode={2795}, 
            state={NSW},
            country={Australia}}

\affiliation[inst2]{organization={Department of Computer Science},
            addressline={St. Francis Xavier University}, 
            city={Antigonish},
            postcode={B2G 2W5}, 
            state={NS},
            country={Canada}}


\begin{abstract}

With the availability of many child sexual abuse (CSA) education apps in different app stores, the demand for an evaluation system has become necessary as little information regarding their evidence-based quality is available.
The objectives of this study are understanding the requirements of a CSA education app, identifying the limitations of existing apps, and providing a guideline for better app design.
An electronic search across three major app stores (Google Play, Apple, and Microsoft) is conducted and the selected apps are rated by three independent raters. Total 191 apps are found and finally, 14 apps are selected for review based on defined inclusion and exclusion criteria.
An app rating scale for CSA education apps is devised by modifying existing scales and used to evaluate the selected 14 apps. Our rating scale evaluates essential features, criteria, and software quality characteristics that are necessary for CSA education apps, and determined their effectiveness for potential use as CSA education programs for children. The internal consistency of the rating scale and the inter and intra-rater reliability among the raters are also calculated. User comments from the app stores are collected and analyzed to understand their expectations and views. After analyzing the feasibility of reviewed apps, CSA app design considerations are proposed that highlight game-based teaching approaches.
Evaluation results showed that most of the reviewed apps are not suitable for being used as CSA education programs. While a few may be able to teach children and parents individually, only the apps ``Child Abuse Prevention" (rate 3.89 out of 5) and ``Orbit Rescue" (rate 3.92 out of 5) could be deemed suitable for a school-based CSA education program. However, all those apps need to be improved both their software qualities and CSA-specific features for being considered as potential CSA education programs.
This study provides the necessary knowledge to developers and individuals regarding essential features and software quality characteristics for designing and developing CSA education apps.

\end{abstract}

\begin{keyword}
Mobile app\sep child sexual abuse\sep education\sep design considerations\sep app rating scale\sep apps review
\end{keyword}

\end{frontmatter}


\section{Introduction}
\label{sec:sample1}
Child sexual abuse (CSA) is a dire public health issue that has adverse effects not only on the victims but also on families, and society~\cite{Singh2014}. CSA as defined by the World Health Organization (WHO), is the engagement of a child in sexual activity that he or she can not completely understand, is incompetent to provide informed approval to, or for which the child is not developmentally fit and cannot give consent, or that breaks the laws or taboos of society~\cite{Csorba2004}. 
CSA includes engaging in sexual activities with a child (whether by demanding or compelling), indecent exposure, child grooming, and also using a child to produce child pornography. This heinous crime is one of the most hidden and unreported forms of violence that affect children’s lives worldwide~\cite{Goldman2000}.

CSA has become extremely prevalent everywhere that it can not be ignored anymore. In recent years, the frequency of sexual abuse targeting children has been growing unwaveringly~\cite{WorldHealthOrganization2020}. Worldwide, one in every ten children is sexually exploited before their 18th birthday~\cite{Darknesstolight}. That means about one in seven girls and one in twenty-five boys will face sexual abuse before they turn eighteen. However, barely one-third of CSA occurrences are identified and even less are reported~\cite{Singh2014}. People who experience sexual abuse in childhood suffer throughout their entire life~\cite{Cho2015}. They face the risk of serious mental disorders, suicide, drug misuse, eating disorders, health problems, and displaying violent behavior. Therefore, teaching both children and adults CSA education has become indispensable. When adults are properly educated regarding the horrors of CSA, they will be more inclined to teach their children the prevention rules. There is no substitute for developing effective CSA educational programs that can teach children both in school settings and via mobile application.

To combat the sexual exploitation of children, various prevention programs have been adopted universally. Despite having these CSA education programs and laws for protecting children, the horrors of CSA keep on growing~\cite{Sanderson2004}. Every day around the world thousands of children are added to this list of being traumatized forever or losing their lives~\cite{Daray2016}. It is evident from the statistics that the laws alone can not keep children out of the perpetrator's grasp ~\cite{Cullen2020}. Most of the prevention programs require children to have regular face-to-face contact with the teachers. The sexual abuse prevention programs that are available now show variable
efficacies due to inconsistencies in the teachings and duration of the programs~\cite{He2001}. Also, they are out of reach for most children who are on the lower end of the economic scale~\cite{Holloway2018}. Even though Parents and caregivers play a huge role in protecting children, school-based programs hardly ever include parents~\cite{Rudolph2018}. Therefore, the obligation exists to design programs that efficiently teach both children and adults about CSA and its prevention.


Recently, several apps (e.g., ``Orbit Rescue", ``Elements of Child Sexual Abuse", ``Game on POSCO", ``Stop the Groomer", and  ``Helpio") have been developed for making CSA education available to all children. Reporting sexual abuse, providing medical information, legal help, and teaching children the rules of safety are the common features of CSA education apps. Some of these apps (e.g., ``iSafe English', ``Game on POSCO", ``Orbit Rescue") use game-based learning, serious games, and others (e.g., ``KidzLive", ``Stop the Groomer", ``The Ceceyara App") use standard informational learning. Good thing is, few of them solely focus on teaching parents and caregivers (e.g,~\cite{BalSurakshaApp}). CSA education apps also provide professional training for social workers and child care professionals~\cite{ElementsApp}. However, it is disputable if these apps are trustworthy as sexual abuse prevention education method~\cite{He2001}. Even though there is a large number of apps claiming to teach prevention, they should be rigorously evaluated before labeling them as reliable. 
To the best of our knowledge, no studies have reviewed or evaluated the currently available apps regarding CSA  to determine their effectiveness in educating children about sexual abuse prevention.

For an app to be competent in teaching children it needs to have some key features such as appropriate content for the target age group, including adults, focusing on developing a wholesome self-concept of players, making the learning environment appealing and relatable for all players, and evaluating children's knowledge~\cite{Stieler-Hunt2014}. Furthermore, the apps need to have basic software quality characteristics for them to be user-friendly and helpful~\cite{Noei2017}. So, it remains a question whether they are actually useful in educating children and parents about CSA.

In this paper, a systematic analysis of the CSA education-related apps has been carried out. A keyword-based search of over 3.5 million apps from three major app stores (e.g., Google Play Store, Apple App Store, and Microsoft Store) was conducted to find the apps that relate to CSA education. 191 apps were found from the electronic search and 14 apps that matched our study criteria were selected for review. Then existing app rating scales were modified to devise one that will be suitable for qualitative and quantitative analysis of our finally selected apps. The internal consistency of the modified rating system shows that this rating scale has high internal consistency. And the score of the inter-rater \& intra-rater reliability among the raters justified the reliability of the study. This study was motivated by the accelerated increase of such CSA education apps in the app stores, along with growing interest among parents about their efficacy. In particular, this research has made the following four major contributions:

 
\begin {itemize}
\item We have performed a systematic review of the existing CSA education apps available in the three major mobile app stores (i.e., Google Play Store, Apple App Store, and Microsoft Store). 

\item We have adopted and extended the existing mobile app rating scales for devising an app rating scale specifically suited for evaluating CSA education apps.

\item We have evaluated the selected apps using our CSA education app rating scale and highlighted their design limitations.

\item Our study also provides design guidelines for developers regarding the aspects which they need to consider while creating an effective CSA education app. Moreover, individuals can also gain valuable information regarding the important features a CSA education app must-have.
\end {itemize}

The rest of this paper is organized as follows. The app search \& selection process and the proposed CSA app rating scale are described in Section~\ref{method}. In Section~\ref{results}, the results found by analyzing the apps, the inter-rater \& intra-rater consistency, and the internal reliability of our rating scale are presented. Design limitations of existing apps and future design considerations for developers and individuals are discussed in Section~\ref{findings}. Finally, Section~\ref{conclusion} concludes the paper.

\section{Methodology}\label{method}

In this section, the overall methodology for reviewing the CSA-related apps has been discussed. The search and selection process of the CSA-related apps has been explained along with the modifications that were done for developing the scale that was used for rating CSA education apps. Also, different sub-scales of the modified rating scale have been discussed in detail.

\subsection{App Search Procedure}

This research includes the apps found in the three mobile app stores: Apple App Store, Google Play Store, and Microsoft Store, which are the most popular mobile application stores presently. The search was conducted between August 2020 and September 2020. A keyword-based search process was employed in Apple App Store, Google Play Store, and Microsoft Store following similar approaches used in previous studies~\cite{Rivera2016}. The Preferred Reporting Items for Systematic reviews and Meta-Analysis or PRISMA guidelines were also followed~\cite{Tricco2018}. Following PRISMA guidelines ensure transparency and clarity of reporting and permits other researchers to replicate the search process. Examining these three stores guaranteed that both Android and iOS apps would be included in the study. The keywords were chosen carefully by analyzing the names of some popular CSA related apps 
, so that if the same set of keywords are used at the same given time point and from the same location, the search will produce the same result~\cite{Stawarz2015}. The keywords that were used are: ``child sexual abuse", ``sexual abuse prevention", ``sexual abuse", ``sexual offence", ``abuse prevention", ``child abuse".

The investigators performed the searching, screening, and final inclusion process collaboratively. For minimizing differences and ensuring consistency, all three app stores were searched using the same enlisted keywords. Three investigators independently performed the exact search applying the same keywords several times and assembled the final inclusion list of CSA education apps. Each investigator maintained a separate list of apps for every app store they searched. They used the inclusion and exclusion criteria (discussed in section~\ref{App Selection Process}) for screening the apps. Every investigator used their smartphone for conducting this search and selection process. The separate lists kept by the investigators were merged for producing the final app list to be reviewed and analyzed for this study. Conflicts between the lists were mutually resolved through discussion among all investigators.

%

\subsection{Raters}
The three raters chosen for the analysis of the CSA-related apps include two final-year Bachelor of Computer Science students with two years’ mobile app development experience, a final-year Bachelor student with extensive user interface design experience. Also, two computer science graduates with two years of experience in HCI and mobile application development research rated one app (Orbit Rescue) for calculating internal consistency. 

The raters separately rated all the apps from the final app list produced by the investigators. Their responses were recorded in a response form (Google Forms) and respective rater response data was extracted from the spreadsheet attached to the form.

\subsection{App selection process}\label{App Selection Process}

\begin{figure}[htbp]
\centering
\includegraphics[width=.85\textwidth]{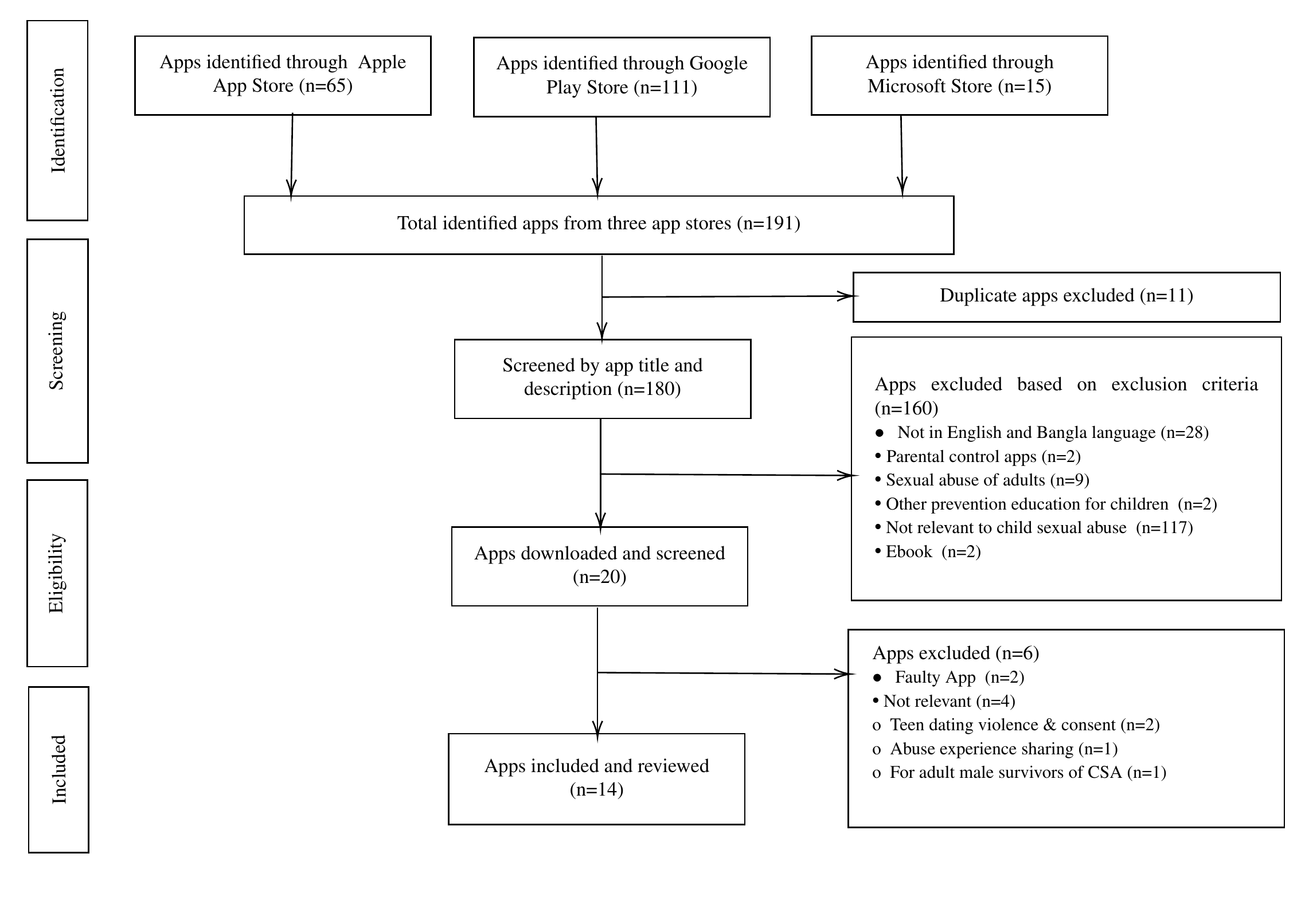}
\caption{PRISMA diagram of the method}
\label{fig: Flow diagram for study method}
\end{figure}

The complete process of app searching, screening, and selection is depicted in the PRISMA diagram in Figure~\ref{fig: Flow diagram for study method}.

The keyword-based search strategy employed on three different app stores during the search process attained 191 apps. After this, the apps were selected based on their app store description. This step served as a primary screening stage. For example, if the app description suggested that it includes CSA-related information we included it in the study. For further curation, the following inclusion criteria have considered: (1) Apps that teach children CSA prevention, (2) Apps that teach parents about CSA, (3) Apps that provide professional training for adults on how to prevent CSA, and (4) Apps that provide legal and medical help to CSA victims. These inclusion standards had been determined for the selection of apps so that the included apps would satisfy our study criteria.

 In this phase, 11 apps were excluded because they had identical apps from the same developer or publisher in multiple app stores (a duplicated app), and therefore were excluded from one platform. Before exclusion, these apps were examined on respective platforms (Android, iOS, and Windows) to check whether they have identical features. The remaining 180 apps were screened by app title and description. From the primarily screened apps, 160 were excluded for one or several of the following reasons: (1) apps whose language was not English or Bangla as these are the languages authors are accustomed to, (2) apps designed for sexual abuse prevention of adults, (3) parental control apps for tracking children’s online activity, (4) apps that teach children other safety practices (e.g., fire safety, road-crossing safety, etc.) rather sexual abuse and were selected because of misleading descriptions, and (5) Apps that were categorized as ``e-books". 

For secondary screening, the remaining 20 apps were downloaded on the rater's smartphone device. And fully assessed by three raters individually.  The apps that didn't satisfy the predefined inclusion/exclusion criteria, were excluded at this stage. At this stage, six apps were excluded because either they were faulty or had content that is not the focus of this study (e.g., experience sharing, adult male survivors of CSA, etc.). Finally, the 14 apps selected in this study are evaluated. The details of the 14 apps (name, developer, country of origin and language, intended platforms) are presented in Table~\ref{table2List of child sexual abuse prevention mobile apps}. All 14 apps were from the Google Play Store and Apple App Store. There is no CSA-related app reviewed from Microsoft Store. The majority of the apps are from the United States and India. From 14 apps only one app (``Shishuk jouno hoirani theke bachate koronio") is from Bangladesh and in the Bangla language. The ``Child Abuse Prevention" app was available in two languages: English and Vietnamese.

\begin{table}[!htb]
\caption{List of CSA education mobile apps included in quantitative and qualitative analysis}
{\small
\begin{center}
\begin{tabular}{p{5cm}p{4cm}p{3.7cm}p{2.4cm}}
\hline
App Name & Country of Origin & Language & Platform\\ \hline
\hline
Orbit Rescue & Australia & English & Android \& iOS \\ 
Elements of Child Sexual Abuse & United States & English & Android\\ 
Bal Suraksha & India & English & Android\\ 
{Shishuk jouno hoirani theke bachate koronio}&  Bangladesh & Bangla & Android\\ 
Stewards of Children scalekit & United States & English & Android \& iOS \\ 
Game on POCSO
& India & English & Android \\ 
Stop the Groomer & United States & English & Android \& iOS \\ 
ChildAbuseInfo & United States & English & Android \& iOS \\ 
iSafe English & India & English & Android \& iOS \\ 
Feel Safe  & Australia & English & Android \& iOS \\ 
Helpio  & Nigeria & English & Android \& iOS \\ 
Child Abuse Prevention  & Vietnam & English \& Vietnamise & iOS\\ 
KidzLive  & Singapore & English & iOS \\ 
The Ceceyara App  & Nigeria & English & iOS \\ 
\hline

\hline
\end{tabular}
\label{table2List of child sexual abuse prevention mobile apps}
\end{center}
}
\end{table}

\subsection{CSA apps rating scale 
}

We hypothesized that for properly evaluating an app's efficiency in teaching CSA and spreading awareness among the users, the app should include specific features and fulfill certain criteria. As there are many apps regarding CSA on the app stores, a standardized rating scale would help evaluate the viability of these apps. To learn significant criteria and standards for evaluating the usefulness of mobile apps, an extensive review of existing mobile app rating scales was conducted. After analyzing the existing app rating scales~\cite{Faux2016, Poon2015, Friesen2013, Vos-Draper2013, Kabir2021}, we concluded that each of them is suited for distinct categories of apps. We focused on essential software quality attributes such as usability, reliability, functionality, and performance \& efficiency after analyzing existing researches on software quality evaluation. 

Guidelines for assessing the usability of CSA education apps have not been set by any previous research. As this is a very sensitive and extremely important issue, we wanted to make sure that the app rating scale we develop is suitable for assessing this category of apps. For this, we took insight from the existing rating scales such as the mobile app rating scale (MARS)~\cite{Stoyanov2015}, end-user version of the MARS (uMARS)~\cite{Stoyanov2016}, and app rating scale for foot measurement apps (FootMARS)~\cite{Kabir2021}, and modified these for suiting the required evaluation of CSA education apps. We especially adopted distinctive categorical regions from the FootMARS rating scale ~\cite{Kabir2021} and introduced an extended variant of the mobile app rating scale for rating CSA education apps. This rating scale includes modifications that we hypothesized to be more important for a CSA education app. It considers specific items that were discovered during the analysis of the selected CSA education apps and previous research on the effectiveness of CSA education programs. 

The fundamental domains that are crucial for the assessment of CSA education apps can be summarized as app classification, aesthetics, general features, performance and efficiency, usability, CSA education-specific functionality, transparency, subjective quality, and the app’s perceived impacts on the user. 

The app metadata and app classification category collects descriptive information about the app. These sub-scale items were omitted from quantitative measurement. The app quality assessment is centered around the domains aesthetics, general features, performance and efficiency, usability, application-specific functionality, transparency, subjective quality, and the app’s perceived impacts.

We have prepared a questionnaire containing a total of seventy-one questions for the app quality assessment based on all app quality domains. Fifty-one of these questions are based on the Likert scale. The questions that used the Likert scale to obtain the user's level of agreement or disagreement were all positively scored. The most agreement means score five and the most disagreement score is one. There are some binary scale questions. However, these questions are later converted into a Likert scale for efficient calculation. In situations where a question may not be suitable for all apps, the option ``Not Applicable" was included. Also, in some cases all the information could not be accessed regarding an app. The option ``Not Accessible" was provided for such cases. The options ``Not Applicable" and ``Not Accessible" were disregarded during the final calculation for app evaluation. The overall score of this rating scale is achieved by adding the mean of individual app quality domains.

\subsection{App metadata}
Generic information about the apps is extracted from respective app stores under the app metadata category. App metadata incorporates information such as app name, developer information, store URL, platform, number of downloads, star rating, and country of origin. As this information is excluded from the evaluation, it does not have an impact on the rating scale. Three investigators collected the app metadata from each app store via filling a Google Form and the data were automatically stored on a Google Sheet. Also, these data were cross-checked after the extraction by all the investigators.

\subsection{App classification}
After analyzing the apps, we have categorized them into the following subcategories (see Table~\ref{table4Category-wise App name}): (1) CSA education using games, (2) Information guide for children, (3) Professional training for CSA for child care officials, (4) Information guide for parents, teachers, and caregivers.

CSA education through games apps has incorporated either game-based learning or gamification to engage children and make them aware of the risks. Some apps also teach prevention rules via serious games. Prior research shows that the most effective way of teaching children is via game or game-based environments~\cite{Stieler-Hunt2014, Moon2017}. The second category apps provide a CSA prevention information guide using information portrayed in charts or tables or paragraphs. The apps in the third category are targeted towards adults and provide online professional training for child care officials. Category four includes the apps that guide parents, teachers, and caregivers as to how they will protect their children and handle difficult situations. CSA education apps must provide suitable content for the target age groups (e.g., adolescents, young, and adults). The target age group for each of the apps has been listed in Table~\ref{table4Category-wise App name}.

\begin{table}[htbp]
\centering
\caption{Category-wise app name and target age group}
\label{table4Category-wise App name}
{\small

\begin{tabular}{p{4cm}p{6cm}p{5.5cm}}
\hline
Category& App Name & Target age group \\ 
\hline
\hline
\multirow{6}{4.5cm}{CSA education using games}& Orbit Rescue & Children under age 12 \\ 
& Game on POCSO English Version Online & General\\ 
 & Stop the Groomer & Children (under 12) \& Adolescents (13-17) \\ & iSafe English& Children (under 12)\\ 
 & Feel Safe& Adolescents (13-17)\\ 
& Child Abuse Prevention & Children (under 12)\\ \hline
\multirow{2}{4.5cm}{Information guide for children} & Helpio& General, Above age 12 \\ 
 & The Ceceyara App & General \\ \hline
\multirow{2}{4.5cm}{Professional training for CSA prevention for child care officials} & Elements of CSA & Adults (Child care professionals)\\ 
& ChildAbuseInfo & (Child care professionals)\\\hline
\multirow{6}{4.5cm}{Information guide for parents, teachers, and caregivers}& Bal Suraksha & Adults (Parents \& caregivers)\\ 
 & Shishuk jouno hoirani theke bachate koronio & Adults (Parents \& caregivers)\\ 
& Stewards of Children scalekit & Adults (Parents \& caregivers)\\ 
 & Helpio & General (Parents \& caregivers)\\ 
 & KidzLive & Adults (Parents \& caregivers)\\ 
& The Ceceyara App & General (Parents \& caregivers)\\ 
\hline

\hline
\end{tabular}
}

\end{table}

\subsection{Aesthetics}
The mobile application market is extensive at present. The internet is overrun with thousands of similar categories of apps with comparable purposes and outcomes. But the question is what makes an app better than others. A blend of innovative ideas, compact engineering, and intelligent design is the answer~\cite{Speckyboy2020}. But there’s something else that plays an indispensable role in ascertaining an app’s success, which is the visual appeal of the app. Precise layout and organized user interface elements in an app can draw the line between an app’s success and downfall in today’s competitive world. This quality is especially important in CSA education apps. One of the main aspects of determining the aesthetics quality of an application is that it is visually appealing to the users. As the main target users of the CSA education apps would be children, the apps need to be intriguing to the children. Furthermore, the layout and organization of the user interface have to be clean, precise, and child-friendly. If any of these characteristics is missing, then the apps would not be able to reach their perceived goal. To measure the aesthetic quality of an app, the raters considered three factors: (1) the layout consistency and readability of the apps, (2) Content resolution, and (3) the visual appeal of the app.

\subsection{General features}

For educating children about CSA efficiently, the apps must contain some specific features such as addressing grooming, having age and gender-specific knowledge, having information for parents, etc. Apart from these apps must have some other general features such as social sharing and data exporting features. These features enable the users to store what they have learned for future reference and share the knowledge with others.  The authentication feature removes the user’s dependencies on a particular mobile device for using the app. This feature is enabled by storing data with individual user’s credentials on the cloud with additional security features. A study found that content customization options and the presence of more visual information in apps improve its user value~\cite{Stieler-Hunt2014}. The on-board tutorial helps to use the app properly which is an important feature, especially for apps that are developed for children. Regular notification from a learning app may increase the usage frequency of that app resulting in better learning outcomes~\cite{Hamari2016}. Some apps provide premium subscription packages which may provide a better user experience, so, this feature was also included in the rating scale. The absence of these qualities was counted as low points while rating the apps.  


\subsection{Performance and efficiency}
One of the factors that contribute most to the functionality of an app is its performance rate and how efficiently it can run on the user’s device. However, the features evaluated for determining performance scores may vary on different devices. The same app may perform differently across different mobile devices as a subject to CPU performance, total memory usage, total battery life impact, the level of device heating, etc. The raters used the data found from App settings (Android) or auxiliary software (iOS) at 15-minute intervals for tracking the battery usage of the apps. An ideal app must have the ability to work efficiently on all of the devices. The apps should be lightweight and work considerably faster since children have minimal patience comparing to adults. Also, the apps that provide online courses and reporting require increased processing power than others.
Raters thus tracked the following points for evaluating the performance efficiency of the apps: (1) how much time the app takes for responding, (2) whether the features and components work accurately, (3) whether the app crashes frequently, (4) how much memory storage the app takes, and (5) whether there were any noticeable changes in device temperature during usage.

\subsection{Usability}
One of the most vital steps of app development is usability. It performs an important role in shaping users' valuable experiences. This can be defined as the property that evaluates how easily a system interface can be used. Previous works involving the human-computer interaction (HCI) and user-centered design showed that non-intuitive gestures and poor organization of app elements are the usability defects that drive users to abandon some mobile apps and search for better alternatives~\cite{Torous2018}. Usability testing of an app is crucial in deciding whether the app has an adequate quality to draw the attention of its target user groups. For an app to be successful, it must be intuitive, and the user should be able to gain some level of familiarity with the interface in a very short time. Navigation and ease-of-use features come into play to ensure this. Higher ease of use increases the probability of an app remains installed on the user's phone. The usability of an app in real life fluctuates considerably compared to that in laboratory settings due to the diversity of user behavior and the user experience~\cite{Kekalainen2005}. This implies that the usability testing of an app is a very crucial part of app development. By analyzing previous app rating systems~\cite{Stoyanov2015, Stoyanov2016}, we evaluate an app’s usability as ``good" if it provides the following qualities: (1) the app can be operated with ease, (2) the app can be navigated without any interruption, (3) the gestural designs and screen links (e.g., buttons, arrows, navigation panels, etc.) work uniformly across the whole app, (4) the app presents an interactive experience. These requirements were examined by the raters when rating the usability criteria of an app.

\subsection{App specific functionality}
The availability of certain features is one of the fundamentals for the apps that are created for expanding awareness and prevention education about CSA. However, having more properties does not imply that the app will be capable of achieving its target goal. CSA is a terrible and extremely sensitive issue, hence any application concerning this issue will only be beneficial when it will be able to engage the children's interest and efficiently train them in the safety rules. Also, the app must be able to spread knowledge and awareness among parents and caregivers on how to protect their children~\cite{Walsh2012}. 

Apps regarding CSA can be divided firstly into two categories: the first category includes the apps that target children's education and the second type of apps are for parents and caregivers. Apps created for children also vary in factors regarding the approach followed for teaching children. Some apps utilize game environments or components as a method of education. Other apps use information and quizzes for teaching prevention rules. Several CSA education apps provide professional training for social workers and child care professionals. There are also apps for spreading awareness among parents about the dangers of CSA. Few apps have provided education for both adults and children. The common features of CSA education apps were identified by reviewing the existing apps. 

Therefore, after taking into account the important guidelines for CSA education apps, we constituted the app-specific functionality rating for CSA education apps. A major part of CSA education is teaching children the prevention rules that will help them stay safe from abuse~\cite{Kang2020}. It is also important to educate parents and children regarding the risk of grooming techniques used by perpetrators to abuse children~\cite{Bennett2020}. Previous research regarding the efficacy of CSA education points to the fact that teaching methods such as game-based learning, gamification or serious games work better in teaching children prevention education~\cite{Stieler-Hunt2014}. And evaluating children's knowledge after the lessons is necessary to identify the effectiveness of such apps. Research also shows the importance of providing age and gender-specific CSA education to children~\cite{He2001, Scholes2012}. Parents play a crucial role in protecting children from abuse. Hence, their involvement in the teaching process is also necessary~\cite{Walsh2012}. It is also important for apps to provide information regarding help centers for medical and legal aid to sexually abused children. Hence, we included these two points in our CSA education-specific sub-scale.

To be considered useful in preventing CSA, an app must include these subsequent features: (1) prevention education for children, (2) addresses the risk of grooming, (3) uses game-based learning/ serious games/ gamification, (4) evaluates children's knowledge, (5) provides age-specific education, (6) provides gender-specific education, (7) facilitates parent involvement, (8) provides guidance to seek legal help, (9) provides medical/counseling help, (10) applicable for physically/mentally challenged children, (11) provides abuse reporting facility.

\subsection{Transparency}
Privacy is an increasing concern in the digitally connected world. Mobile apps require user's social and personal information for proper functioning. However, the ability of consumers to make informed choices regarding their privacy is difficult at present in the mobile application marketplace. Businesses that capitalize on these personalized services make this choice even harder. Often, apps trade user's private data without awareness of the individual~\cite{chang2020}. App users fall into these traps due to insufficient knowledge regarding mobile application privacy policy. Users must be conscious while consenting to their private data being accessed by the apps. They should make a habit of properly reading the privacy guidelines before clicking the `Accept' button in any circumstance. 

Data protection and regulation rules must be followed strictly by the apps. Furthermore, they should explicitly explain how and why they are collecting users' data. In the case of CSA education apps, the privacy guidelines should be followed strictly. Evaluation of transparency is a significant part of the app rating. With the proper and clear information, a user can make an informed decision before downloading the app by ascertaining the authenticity of the app.
To assess the transparency criteria of the selected apps, the following points were considered: (1) whether a common alert is provided to take users consent before obtaining their personal information, location information, and/or private data, (2) evaluating the accuracy of the information given on the app store description, (3) determining the authenticity of the publisher or developer and the app source, (4) whether the app is feasible to achieve desired goals as claimed by the developer.

\subsection{Subjective quality}
The subjective quality of an app is the perspective of a user of the app. This quality can be measured by the user's app ratings and comments about the app. A lot of information regarding the performance of an app can be found from the reviews in the app store as the users these days often comment about their user experience in detail. Both the flaws and the unique features of an app can be discovered from user reviews. These comments are somewhat helpful during the app review phase. So subjective quality analysis can be a useful criterion for app evaluation considering its benefits. In this study, a similar approach was used for the subjective quality section. The app raters evaluated subjective quality by answering questions about the degree of satisfaction of use, potential frequency of use in the future, overall app rating, and how likely they were to pay for the apps.  


\subsection{Perceived impact of app on users}
The success of an app can be determined by the impact it has on its users. Continuous discoveries in the branches of computer science and mobile apps have assisted thousands of users in finding useful and sometimes life-saving information, including a large number of apps that are designed to provide different prevention education (e.g., sexual abuse prevention, drug abuse prevention) to users~\cite{Tait2018, Jones2020}. In apps regarding prevention education, the main goal is to increase awareness about crime or addiction and provide necessary information about the prevention of this problem. Additionally, such apps may also provide intervention methods and guidance helpful in decreasing the user’s negligence towards the specific problem and increasing help-seeking behaviors targeting solutions to the problems. However, according to the study by~\cite{Milne-Ives2020}, most health-oriented and prevention education apps yield little to no evidence of effectiveness in cases of user outcomes and behavioral changes. 

For determining the usefulness of the CSA education apps, these apps must be evaluated by their effectiveness regarding changing the attitude of the users towards CSA education and spreading awareness about CSA~\cite{Milne-Ives2020}. App store comments and ratings can be used for evaluating the perceived impact an app has on its users. From the user reviews, we can get an idea about what kind of effect the app has had on its users. 
The actual impact of an app on its users is can not be directly quantified. Therefore, the apps were rated on the following aspects to assess how much an app was able to make an impact on its users: (1) whether the app contained information for raising awareness regarding CSA, (2) how informative the app is considering CSA education and dangers of CSA, (3) whether the information or learnings of the app will improve users attitude about CSA education's necessity, (4) whether the app will induce further help-seeking behavior among users.

\section{Results}\label{results}
In this section, an overall assessment of the reviewed 14 apps has been presented. Also, a comparison between the app store rating and our rating is given. The app-specific criteria have been discussed in detail. Lastly, we analyzed the app store user comments of the apps for understanding the good and bad factors of existing CSA-related apps.

\subsection{Internal consistency of modified rating scale}
Cronbach’s alpha is the most popular means of calculating internal consistency. It denotes how well a test measures, and how consistent the items of a scale are~\cite{Cronbach1951}. This method has been used for calculating the internal consistencies of the sub-scales of our modified rating scale: aesthetics, performance and efficiency, usability, transparency, subjective quality, and perceived impacts on users (see Table~\ref{table5Internal consistency}). For determining internal consistency, one of the CSA education apps ``Orbit Rescue" was selected. Additionally, the internal consistency of the entire modified rating scale was estimated. Cronbach’s alpha reliability coefficient usually ranges between 0 and 1. The closer Cronbach’s alpha coefficient is to 1.0 the higher the internal consistency of the scale~\cite{Cronbach1951}. Generally, a coefficient value greater than 0.75 is considered acceptable. However, a coefficient value greater than 0.5 is also accepted if the internal consistency is calculated for limited items. For five independent raters, the internal consistency of all sub-scale items was moderate to high. The overall internal consistency of the modified rating scale was high at alpha = .93 which is considered excellent according to prior studies~\cite{Gliem2003}. The sub-scale alphas were also in the range of good-excellent as can be seen in Table~\ref{table5Internal consistency}. The alpha values indicate that all the items in our modified rating scale had a high level of internal consistency. For getting positive values of Cronbach's alpha, we made sure that all questions of the scale were coded in the same way. No positively and negatively worded questions were mixed. Reverse coding was done for the negatively worded questions.

\begin{table}[!htbp]
\caption{Internal consistency of the sub-scales}
\begin{center}
\begin{tabular}{ll}
\hline
CSA App Rating Scale& Internal Consistency \\
\hline
\hline
Aesthetics & 0.92 (95\% CI 0.66-0.99)\\ 
Performance and efficiency & 0.94 (95\% CI 0.80-0.99)\\ 
Usability & 0.81 (95\% CI 0.22-0.97)\\ 
Transparency& 0.84 (95\% CI 0.33-0.98)\\ 
Subjective quality& 0.70 (95\% CI 0.52-0.97)\\ 
Perceived impact of app on users & 0.93 (95\% CI 0.77-0.99)\\ 
\hline

\hline
\end{tabular}
\label{table5Internal consistency}
\end{center} 
\end{table}

\subsection{Inter-rater and intra-rater reliability of the modified rating scale}
The evaluation of inter-rater reliability provides a way of quantifying the level of agreement between two or more raters who make independent ratings about the features of a set of subjects~\cite{Sawa2007}. The significance of inter-rater reliability depicts that the degree to which the data collected in the study consists of correct descriptions of the variables measured. When a scale is judged by several raters, inter-rater reliability is significant to evaluate if their rating is correlated. To find out the appropriate method for calculating inter-rater reliability two things should be considered. Firstly, it is required to know if all items included in a study have been rated by several raters, or if only a single subset is rated by multiple raters. Secondly, it must be determined if the same set of raters will rate the items or whether separate items are rated by separate subsets of raters. In our study, all apps were rated by the same three raters. We used the intra-class correlations (ICC) method to assess inter-rater reliability. The Intra-class Correlation (ICC) is one of the most popularly used statistics for evaluating inter-rater reliability if a study includes two or more raters~\cite{Hallgren2012}. We have used the ICC two-way mixed model since the use of this model is preferred when each app is assessed by each rater, and the raters are fixed but not randomly chosen~\cite{Koo2016}. Depending on the 95\% confidence interim of the ICC estimation, values smaller than 0.5, within 0.5 and 0.75, within 0.75 and 0.9, and higher than 0.90 are suggestive of poor, moderate, good, and excellent reliability, sequentially~\cite{Koo2016}. Independent ratings on the overall scale for our final fourteen CSA education apps exhibited a high level of inter-rater reliability. Our calculated inter-rater agreement is 2-way mixed ICC = 0.97, 95\% CI 0.97-0.98, which can be considered as a great level of rater reliability or agreement~\cite{Hallgren2012}.

Intra-rater reliability is estimated to measure the validity of a test. This is a kind of reliability estimation in which the same evaluation is performed by the same rater on more than one occasion. These different ratings are then compared, generally using correlation. Again, we have used the ICC method for analyzing the intra-rater reliability of our three raters. Three raters rated the final fourteen apps twice in the interval of four weeks. After completion of both ratings, consistency between two ratings of the same rater was calculated to see if their rating is reliable. This was done for all three raters -- rater one showed high reliability between her two ratings (2-way mixed ICC = .97, 95\% CI 0.97-0.98), rater two and three also presented excellent intra-rater reliability of (2-way mixed ICC = 0.96, 95\% CI 0.96-0.97) and (2-way mixed ICC = 0.98, 95\% CI 0.98-0.99).

\subsection{Categorical distribution of CSA education apps}
\begin{figure}[!htbp]
\centering
\resizebox{12cm}{!}{
\begin{tikzpicture}
  \pie[text=legend]{37.5/CSA prevention education using games,37.5/Information guide for parents\, teachers\, and caregivers, 12.5/Information guide for children, 12.5/Professional training for CSA prevention for child care officials}
\end{tikzpicture}}
\caption{Categorical distribution of apps}
\label{fig2Categorical distribution of apps}
\end{figure}
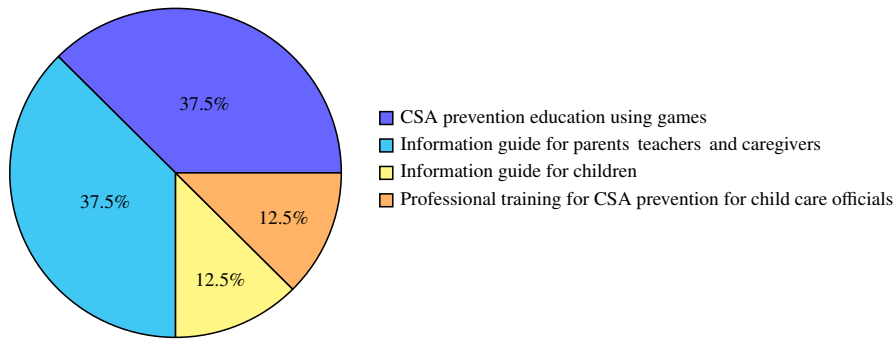

The categorical distribution of all reviewed apps is shown in Figure~\ref{fig2Categorical distribution of apps}. 
It shows that both the ``CSA prevention education using games" and ``Information guide for parents, teachers, and caregivers" categories hold 37.5\% of the total 14 apps. Also, 12.5\% apps are placed in both the ``Information guide for children" and ``Professional training for CSA prevention for child care professionals" categories. Among the 14 apps, two apps are fall in both the ``Information guide for children" and ``Information guides for parents, teachers, and caregivers" category (``Helpio", ``The Cece Yara App" ).
 
It is evident from Figure~\ref{fig2Categorical distribution of apps} and Table~\ref{table4Category-wise App name} that most of our final fourteen apps belong to ``CSA prevention education using games" and ``Information guide for parents, teachers, and caregivers" category. Research shows that game-based learning or learning via serious games or using gamification components in teaching prevention programs are more efficient than traditional rote learning~\cite{Stieler-Hunt2014}. Games help in keeping children engaged and make the learning process enjoyable~\cite{BEHNAMNIA2020}. CSA education using games can help children to learn the safety rules more profoundly~\cite{Jones2020}. Children can learn about the dangers of CSA while playing the game and also get to know how to prevent CSA effectively. Apps in this category use different methods for engaging children such as animated videos where cartoon characters explain the safety rules, mini-games that teach how to overcome difficult situations or interactive evaluation of the knowledge gained from a lesson. 

The other category that has the most apps is the ``Information guides for parents, teachers, and caregivers". These apps spread awareness about CSA and educate parents on how they can protect their children. Parents play a very important role in protecting children from abuse. Hence, apps in this category are extremely significant. Moreover, parents can also learn how to handle disclosure to abuse and how to treat abusers.

\subsection{Evaluation scores}
We calculated the mean and standard deviation of the sub-scale ratings for each app in Table~\ref{table6Evaluation scores}. The scores an app received in different sub-scales were used to calculate the overall mean and standard deviation for that app. From the table, we can see that app that scored the lowest score is named ``Shishu k jouno hoyrani theke bachate koronio". The overall mean score for this app is 2.29 which shows that it is not of great quality and is not effective for spreading prevention information. The reason behind this low score is also visible on the table. This app scored low in all the rating sub-scales and as a result, had an overall low score. It belongs to the category ``Information guide for parents, teachers, and caregivers". On the other hand, the highest mean score is 3.92. The app ``Orbit Rescue" received this score. This app scored more than average in all the rating sub-scales.

\begin{table}[htb]
\caption{Evaluation scores for CSA education apps}
{\small
\begin{center}
\begin{tabular}{p{3.7cm}p{1cm}p{1cm}p{1cm}p{1cm}p{1cm}p{1cm}p{1cm}p{1cm}p{1.8cm}}
\hline
App Name& Aesth-etics & Gener-al & Perfo-rmance & Usabi-lity & Appli-cation & Trans-parency & Subje-ctive & Impact & Mean (St.dev) \\ \hline
\hline
Orbit Rescue& 4.25 & 3.00& 4.00 & 3.75 & 3.57 & 4.25 & 3.67& 4.83 & 3.92 (0.55)\\ 
Elements of Child Sexual Abuse & 3.50& 1.50 & 4.57&3.75& 1.29 & 3.25 & 2.67 &3.50& 3.00 (1.13)\\ 
Bal Suraksha& 3.25 & 1.50 & 4.43 & 3.25 & 2.71 & 3.25 & 3.00 & 4.17 & 3.20 (0.90)\\ 
Shishuk jouno hoirani theke bachate koronio & 2.75& 1.00 & 4.57 & 3.50& 1.86& 1.50 & 1.33& 1.83 & 2.29 (1.22)\\ 
Stewards of Children scalekit& 2.50 & 2.00 & 4.29 & 2.75 & 3.29 & 4.50& 4.00 & 4.67 & 3.50 (1.01)\\ 
Game on POCSO English Version Online & 4.00& 1.50& 4.14 & 3.50& 2.71 & 3.50 & 3.33 & 4.33 & 3.38 (0.92)\\ 
Stop the Groomer& 4.25 & 1.50 & 4.29 & 4.75& 3.29 & 3.50 & 3.67 & 5.00 & 3.78 (1.10) \\ 
ChildAbuseInfo & 3.00 & 1.50 & 4.00 & 2.75 & 3.29 &3.50& 2.33& 4.83 & 3.15 (1.02) \\ 
iSafe English & 2.75 & 2.50 & 3.86 & 2.75& 3.29& 3.00 & 2.00& 3.33 & 2.94 (0.57)\\ 
Feel Safe & 2.75 & 2.00 & 4.00 & 3.00 & 3.00 & 2.75 & 3.67 & 4.17 & 3.17 (0.73)\\ 
Helpio & 3.00 & 2.50 & 4.43& 3.25 & 3.00 & 2.75& 2.33 & 3.67& 3.12 (0.68) \\ 
Child Abuse Prevention& 4.50& 3.00& 4.14& 4.25& 3.00& 3.25& 4.33& 4.67 & 3.89 (0.69)\\ 
KidzLive& 3.25& 1.00& 4.00& 3.00& 2.43& 3.75& 3.00& 3.67& 3.01 (0.96)\\ 
The Ceceyara App& 3.25& 1.00& 3.86& 3.75& 4.14& 3.25& 3.00& 4.00& 3.28 (1.01)\\ 
\hline

\hline
\end{tabular}
\label{table6Evaluation scores}
\end{center}
}
\end{table}
 
\begin{figure}[htbp]
\centering
\resizebox{8cm}{!}{
\begin{tikzpicture}
    \begin{axis}[
        ybar,
        bar width=15pt,
        xtick distance=1,
        ytick={1,1.5,2, 2.5,3,3.5,4,4.5,5},
        symbolic x coords={Aesthetics,General,Performance,Usability,Application,Transparency,Subjective,Impact,Overall},
        ylabel={Rating},  
        xlabel={Sub-scales},
        label style={font=\huge},
        ymin=1,
        ymax=5,
        scaled ticks=false,
        x tick label style={font=\LARGE, rotate=45, anchor=east},
        y tick label style={font=\LARGE},
    ]
        \addplot+ [draw =cyan,
        fill = cyan,
            error bars/.cd,
                y dir=both,
                y explicit relative,
        ] coordinates {
            (Aesthetics,3.36) +- (0,0.05)
            (General,1.82) +- (0,0.05)
            (Performance,4.18) +- (0,0.05)
            (Usability,3.43) +- (0,0.05)
            (Application,2.92) +- (0,0.05)
            (Transparency,3.29) +- (0,0.05)
            (Subjective,3.02) +- (0,0.05)
            (Impact,4.05) +- (0,0.05)
            (Overall,3.26) +- (0,0.05)
        };

        \legend{
        }
    \end{axis}
\end{tikzpicture}}
\caption{Sub-scale specific ratings and overall rating} \label{fig3Overall app rating}
\end{figure}
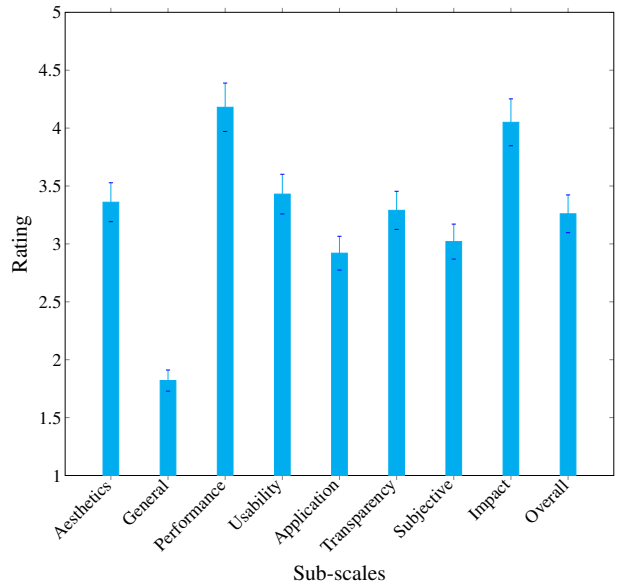

The results in the table show that the domains aesthetics, performance \& efficiency, and usability have higher scores in contrast to other domains of the apps. The general range of aesthetics, performance \& efficiency, and usability stands between 2.50 to 4.50, 3.86 to 4.57, and 2.75 to 4.75 respectively. Among the above-mentioned domains, general app features and application-specific functionality domains received the lowest mean rating (1.82 and 2.92 respectively out of 5). In contrast, the most highly rated domains were app performance \& efficiency and perceived impact of apps on users which received scores of 4.18 and 4.05 out of 5. Other domains that scored more than 3 out of 5 are aesthetics (3.36/5), usability (3.43/5), transparency (3.29/5), and subjective quality (3.02/5). Domain-specific rating scores and the overall app rating are calculated from the data in Table~\ref{table6Evaluation scores} and have been depicted in Figure~\ref{fig3Overall app rating}. We computed the overall app rating (mean score) from the individual mean scores of each app. The overall app rating is 3.26 out of 5.

\subsection{Comparison of store rating and rating scale measured ratings}
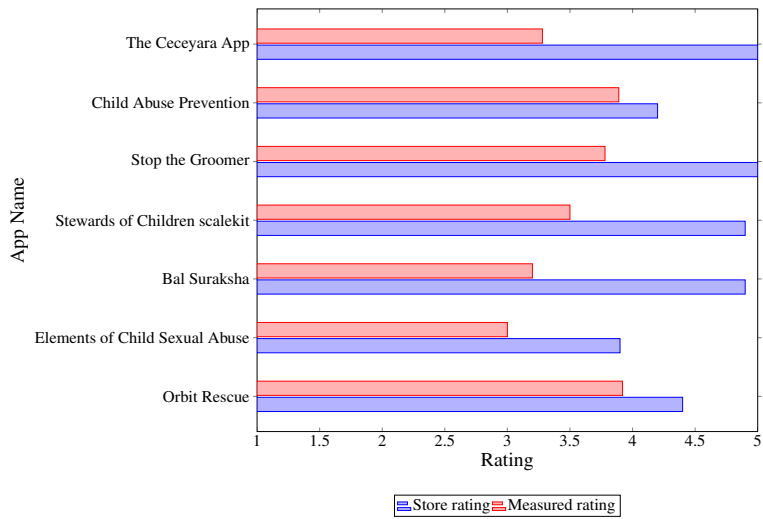
\begin{figure}[htbp]
\centering
\resizebox{10cm}{!}{
\begin{tikzpicture}
\begin{axis}[
    xbar,
    bar width=15pt,
    ylabel={App Name},  
   xlabel={Rating},
   label style={font=\huge},
    legend style={at={(0.5,-0.15),font=\LARGE},
      anchor=north,legend columns=-1},
    symbolic y coords={Orbit Rescue,Elements of Child Sexual Abuse,Bal Suraksha,Stewards of Children scalekit,Stop the Groomer,Child Abuse Prevention,The Ceceyara App},
    ytick=data,
    xtick={1,1.5,2,2.5,3,3.5,4,4.5,5},
    xmin=1,
    xmax=5,
    scaled ticks=false,
    x tick label style={font=\LARGE},
    y tick label style={font=\LARGE},
    %
    ]
\addplot coordinates {(4.40,Orbit Rescue) (3.9,Elements of Child Sexual Abuse) (4.9,Bal Suraksha) (4.9,Stewards of Children scalekit) (5,Stop the Groomer) (4.2,Child Abuse Prevention) (5,The Ceceyara App)};
\addplot coordinates {(3.92,Orbit Rescue) (3,Elements of Child Sexual Abuse) (3.2,Bal Suraksha) (3.5,Stewards of Children scalekit) (3.78,Stop the Groomer) (3.89,Child Abuse Prevention) (3.28,The Ceceyara App)};

\legend{Store rating, Measured rating}
\end{axis}
\end{tikzpicture}}
\caption{Comparison between app store rating and measured rating}
\label{fig4comparison between app store and measured rating scale}
\end{figure}

A comparison between the 14 selected app's app store ratings and the score received from our rating scale has been shown in Figure~\ref{fig4comparison between app store and measured rating scale}. Seven among 14 apps had no rating in the app stores (i.e., ``Shishuk jouno hoirani theke bachate koronio", ``Game on POCSO English Version Online", ``ChildAbuseInfo", ``iSafe English", ``Feel Safe", ``Helpio", ``KidzLive"). Therefore, we excluded them from this comparison. We calculated the standard deviation of the total app store ratings and our rating scale score. The difference between the two standard deviation scores for reviewed apps was 0.08. This deviation is good considering that the score in our rating scale is an aggregated mean of various domains that are necessary for specifying the quality and criteria of CSA education apps. Two of the 14 apps (e.g., ``Stop the Groomer" and ``The Ceceyara App") which had a five-star rating in the app store, didn't get the full score on our modified rating scale and received 3.78 and 3.28 respectively out of 5. The app that scored the highest rating on our scale is called ``Orbit Rescue". This app received 3.92 on our rating scale and 4.40 in the app store. The app that had the second-highest rating according to our scale is ``Child Abuse Prevention". This app had a 4.2 rating in the app store and 3.89 on our rating scale. Overall, the measured rating scores for each app are less than the app store rating value. A possible reason for this difference is that we have determined rating scores by calculating the mean of all sub-scale values.

\subsection{Assessment of measurement criteria in apps}
For evaluating CSA education apps, some specific measurement criteria were defined. The use of game-based components, including information about grooming, gender, and age-specific prevention techniques, providing knowledge for adults, presenting legal \& medical help are some of the key points for measuring the effectiveness of a CSA education app. Our rating scale evaluated the apps based on 11 measurement criteria. These CSA education-specific functionalities and the percentage of apps that have these features are shown in Table~\ref{table7Assessment of Measurement Criteria in Apps}. 

\begin{table}[htbp]
\caption{Assessment of Measurement Criteria in Apps}
{\small
\begin{center}
\begin{tabular}{p{8.4cm}p{2cm}p{2cm}p{2cm}}
\hline
Criteria & \begin{tabular}[c]{@{}l@{}}Google play \\store (n=8)\\ n (\%)\end{tabular} & \begin{tabular}[c]{@{}l@{}}Apple store \\(n=6)\\ n (\%)\end{tabular} & \begin{tabular}[c]{@{}l@{}}Total apps \\(N=14)\\ N (\%)\end{tabular} \\ \hline
\hline
Use game/game based components/gamification & 3 (37.5)& 4 (66.67)& 7 (50) \\
Teach about grooming & 2 (25)& 3 (50)& 5 (35.71)\\ 
Teach children using information & 1 (12.5) & 1 (16.67) & 2 (14.29)\\
Gender specific teaching& 1 (12.5)& 2 (33.33)& 3 (21.43)\\ 
Age specific teaching& 2 (25)& 0 (0)& 2 (14.29)\\ 
Evaluate children's knowledge& 3 (37.5)& 4 (66.67)& 7 (50)\\ 
Parents involvement& 6 (75)& 5 (83.33)& 11 (78.57)\\ 
Provide medical aid& 3 (37.5)& 2 (33.33)& 5 (35.71)\\ 
Provide legal info& 4 (50)& 5 (83.33)& 9 (64.29)\\ 
Provide reporting system& 2 (25)& 4 (66.67)& 6 (42.86)\\ 
Applicable for physically/mentally challenged children & 1 (12.5)& 1 (16.67)& 2 (14.29)\\ 
\hline

\hline
\end{tabular}
\label{table7Assessment of Measurement Criteria in Apps}
\end{center}
}
\end{table}

Table~\ref{table7Assessment of Measurement Criteria in Apps} shows that the `parents involvement' feature is found in 78.57\% (11 out of 14) of apps. The next feature that most apps have is legal information for users. About 64.29\% (9 out of 14) of apps include this feature. To assess the effectiveness of a prevention program, it is important to evaluate how well children have received the teaching. 50\% of the apps (7 out of 14) evaluate children's knowledge regarding CSA and CSA prevention immediately after a lesson. Serious games or gamification were used as a teaching method by 50\% of the apps. Having a reporting system that the users can utilize to report abuse and providing medical aid in case of abuse are also significant criteria that CSA education apps must include. Among the 14 apps, about 42.86\% (6 out of 14) apps had a reporting system, and 35.71\% (5 out of 14) of apps provide medical aid-related information.

Five apps (14.29\%) used information such as charts, graphs, tables for providing prevention information. Only 4 of 14 apps provide knowledge about grooming techniques which is notably an essential aspect of CSA education (Orbit, Stop the groomer, iSafe English, Helpio). Other criteria where the reviewed apps lacked are providing age and gender-specific knowledge and special education for disabled children. Only 2 apps (14.29\%) mention that they can help physically or mentally challenged children in their store description. Also, 3 (21.43\%) apps had gender-specific information and 2 (14.29\%) had age-specific information for children. It is evident from these statistics that gender-specific and age-specific teaching are very rare in reviewed apps. Six of the apps targeted a general age group with no age requirements (42.86\%). These apps were for use of parents, caregivers, and child care professionals. Among all the apps 4 (28.57\%) are targeted for children under age 12. Two (14.29\%) apps have dealt with adolescent users (aged between 13 to 17), which are ``Stop the Groomer" and ``Feel Safe." The app named ``Stop the Groomer" targeted both children and adolescents.

Overall, Among 14 apps, one app (``The Cece Yara App") had seven features from eleven features. Three apps ( ``Stop the Groomer", ``iSafe", ``Helpio" ) included six features. 57.14\% (9 out of 14) of apps have less than half features. Elements of Child Sexual Abuse app had no features and ``Shishu k jouno hoirani theke bachate koranio" app had only one feature.

Most of our reviewed apps are missing several criteria. Most of these apps do not provide CSA lessons in a systematic way which is required for learning efficiently. Children's knowledge regarding prevention should be tested multiple times during the duration of learning~\cite{Stieler-Hunt2014}. Also, any app for CSA and its prevention needs to have trusted adult involvement. Moreover, it must include all of the required measurement features for being effective in rendering prevention education against CSA. In our study, apps that targeted parents and caregivers provide a small amount of information regarding how they can protect their children from abuse, how to deal in case of abuse, where to seek medical or counseling help, etc. Finally, no such app was found during app testing that has focused specifically on children with disabilities.

\subsection{Analysis of user reviews from app store}
The most common way to know whether an app provides all the claimed features is to scrutinize the app store comments left by the users of the apps~\cite{Guzman2018}. Nowadays, all major app stores allow users to critique their hosted apps. These user comments often provide detailed information about the features of the apps, which gives a clear insight to the general users about the apps' performance. These comments point out the shortcomings of these apps as well. Many users rely on the users' reviews before downloading any app. As a consequence, developers and publishers aim to receive good and positive reviews since these reviews act as quality indicators of apps~\cite{Vasa2012}.

Considering the importance of user comment analysis, we collected the users' feedback for the apps from respective app stores while collecting metadata. We divided the user comments into two categories based on the users' rating for the apps: if the rating is 4 stars or more the comment is considered as ``good", otherwise the comment is considered as ``bad". We discovered that most of the apps do not have a sufficient number of feedback from users because the number of downloads for the respective apps was generally low. Two (14.29\%) apps were downloaded only 10+ times. and two (14.29\%) apps were downloaded 100+ times. 3 out of 14 (21.4\%) of apps have been downloaded 1000+ times. And only one app (Elements of child sexual abuse) has been downloaded more than 5000 times. The Apple Store doesn't provide access to the number of downloads for iOS apps (6 out of 14). Moreover, the number of informative reviews was very low as users rarely write informative reasons behind their rating, sometimes they write irrelevant comments~\cite{Vasa2012}. In our study, we discovered that about 35.7\% (5 out of 14) of the apps have no comments and 21.4\% (3 out of 14) have extremely short comments to support the ratings. 

\begin{figure}[!htbp]
\centering
  \begin{subfigure}[b]{.45\textwidth}
    \includegraphics[width=\textwidth]{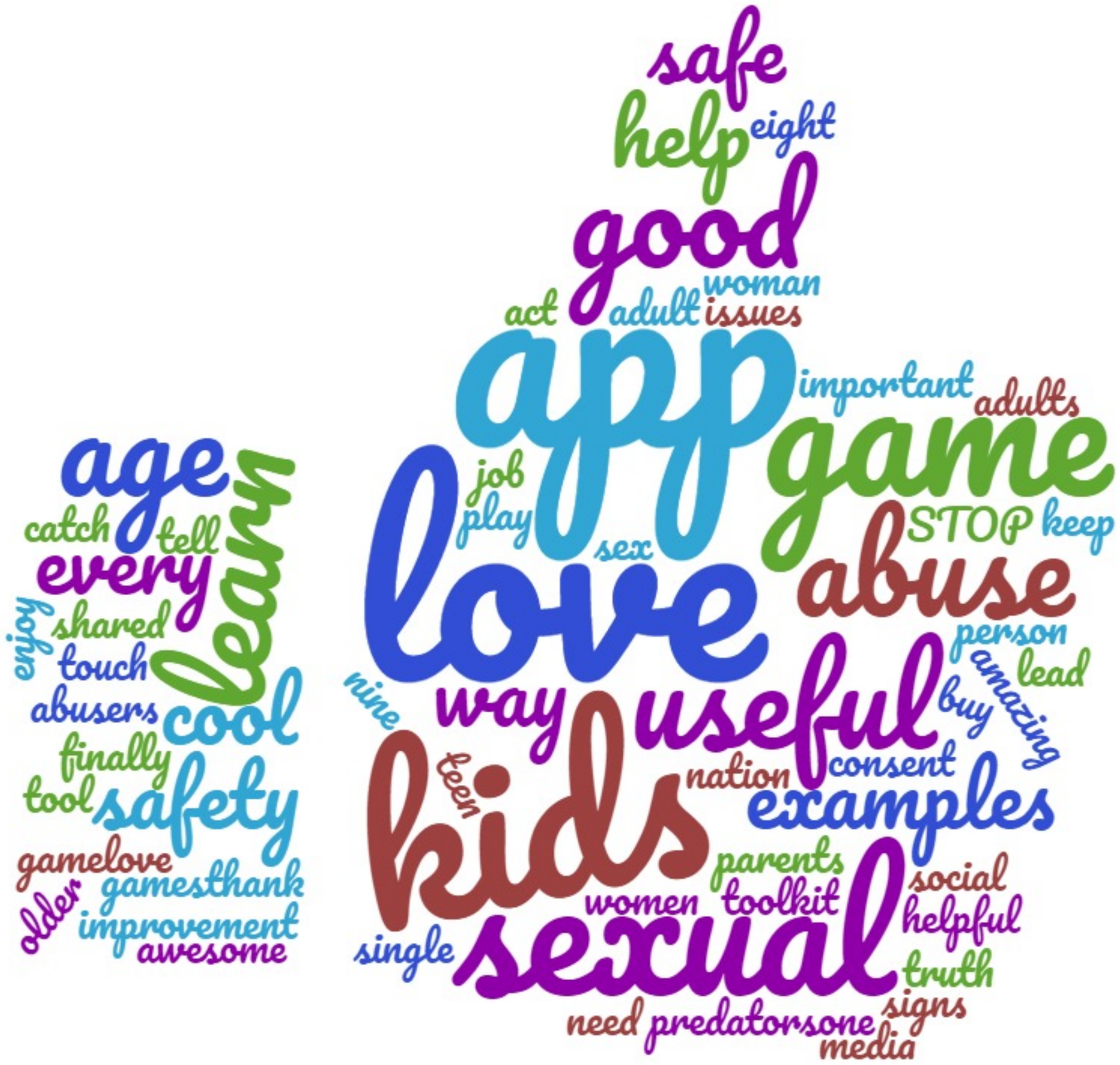}
    \caption{Positive comments}
    \label{fig:good comments}
  \end{subfigure}
  \hspace{0.04\textwidth}
  \begin{subfigure}[b]{0.45\textwidth}
    \includegraphics[width=\textwidth]{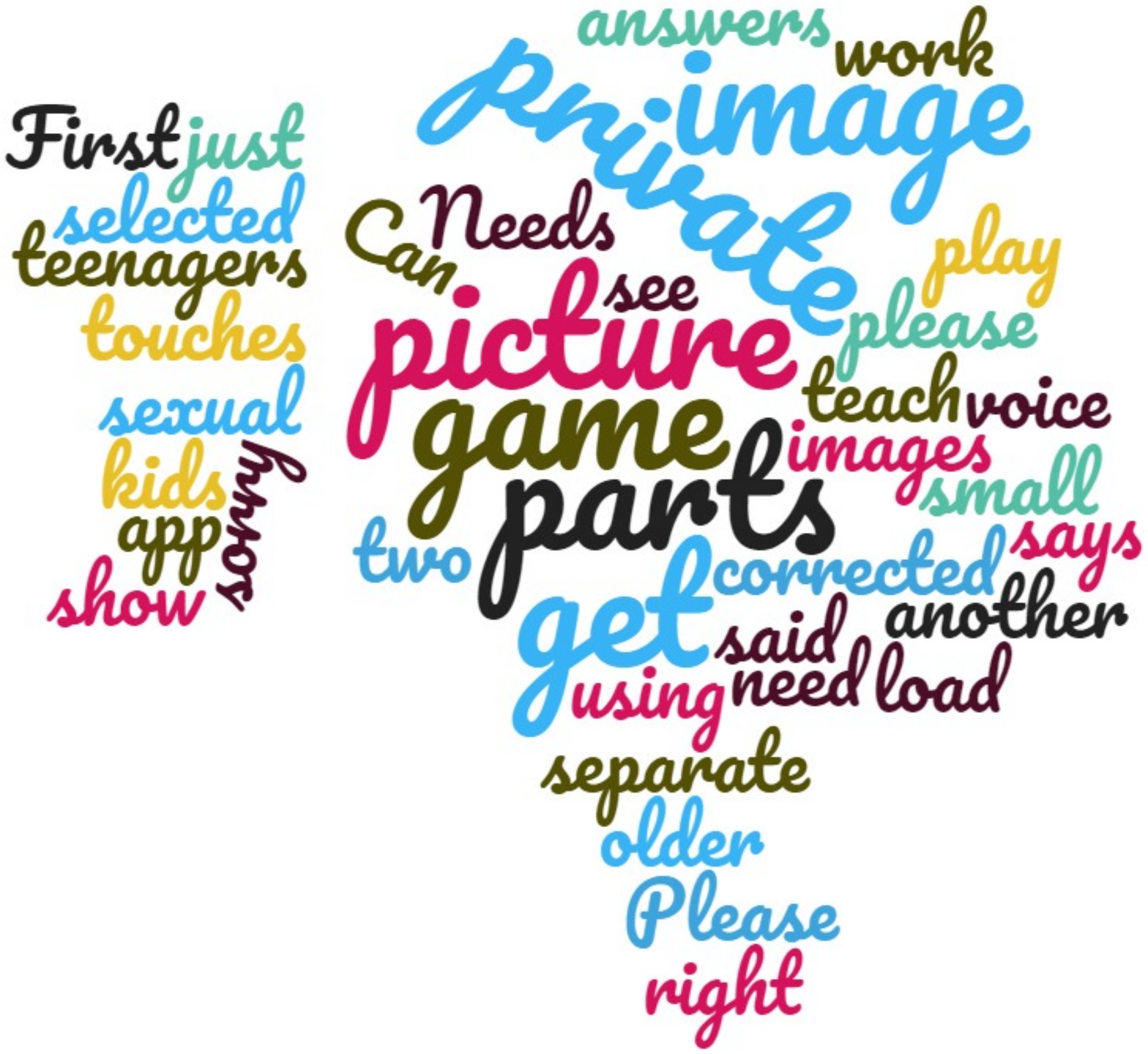}
    \caption{Negative comments}
    \label{fig:bad comments}
  \end{subfigure}
  \caption{Word cloud of good and bad comments}
  \label{Comment comparison}
\end{figure}

To gain better insight from the users' comments, we have used word-cloud for visualizing users' comments. In Figure~\ref{Comment comparison} the positive and negative comments are portrayed. Figure~\ref{fig:good comments} illustrates the words derived from positive comments of the users. The most used words are `app', `love', `kids', `great', `game'. From these words, we can say that users preferred the apps that use a game for teaching children about prevention education. Figure~\ref{fig:bad comments} represents the word cloud for negative comments of the user. The most visible words in this word-cloud are `private', `picture', `parts', `image'. The number of negative comments was low. Therefore, getting an insight from the word cloud is hard. However, the most common words indicate one app that has some technical difficulties in a lesson where the game freezes when selects the private parts. Some apps had received specific complaints in their negative comments for their performance issues such as ``features not working properly", ``app freezing continuously", ``all lessons not being available in the free version". It was perceived from the negative comments that the structural flow of technical aspects was confusing and difficult in some apps which made it difficult for the user to follow.

Only one of the reviewed apps named ``Stop The Groomer" has detailed informative reviews from users. Users referred to this app as both engaging and informative for children. This app focuses on teaching children about the grooming techniques that perpetrators use for abusing them. It includes game-based components where children learn how to stay safe by playing simple games and can implement their knowledge in real-life scenarios. This app has a five-star rating based on the reviews of several users. Good feedback with no negative comments was found in the app titled ``Bal Suraksha" which also obtained good reviews and scored high ratings in the app store. This app belongs to the category of information guide for parents, teachers, and caregivers. Targeted users for this app are parents and it provides information about teaching children safety rules, answering children's questions about sexual abuse, dealing with disclosure of abuse, etc. According to users this app helps spread awareness among adults about CSA education. Another app where user comments were positive and consistent with app store ratings is ``Stewards of Children". This app is targeted towards the adults in a child's life. It provides training regarding CSA to parents and caregivers. Parents can also evaluate their knowledge using this app. Even though this app mostly has good reviews, the rating of this app was low according to our app rating scale.

The app that has the maximum number of positive reviews is ``Orbit Rescue". But the comments are not quite informative in this case. This app uses a serious game for educating children about CSA prevention and it is targeted at children between the age of 8 to 10 years. Users suggested this app as it is an excellent app for teaching safety rules to children. However, it also contains a few negative comments that imply this app might have some technical difficulty. Ratings of our modified scale were consistent with user review and app store rating here. Our rating was also consistent with user reviews in the app called ``Child Abuse Prevention". This app uses a serious game approach and has quite a few reviews in the app store which contains both positive and negative comments. It includes 26 stories that teach children aged between 3 to 10 years about prevention rules so that they can stay safe from abuse. One of the negative aspects that is prevalent in the comments is that this app should be made completely free for wider accessibility.

We have found an interesting observation between the app store rating and user comments. For starters, ``The Cece Yara App" has a five-star rating in the respective app store. However, it doesn't have any user comments. The comment size was low for ``Elements of Child Sexual Abuse" and mostly irrelevant, but the app store rating was high. Also, the apps called ``Helpio", ``Feel Safe", ``Game on POCSO English Version Online" have user reviews, but no star rating in the store. The apps ``Kidzlive", ``iSafe", ``ChildAbuseInfo", ``Shishu ke jouno hoyrani theke bachate koroniyo" had no user comments.

From the analysis of the CSA education apps' user reviews, it can be argued that user comments do not always represent the actual status of the apps. Also, we can resolve that there is a difference between the user comments found in the app store for an app with the rating of our modified scale. One of the reasons for this can be the variation of apps' performance and functionality characteristics observed in different devices as most of the apps we have reviewed performed differently on different devices. This implies that the bulk of the apps currently available in app stores have built-in and device-to-device functioning issues. Hence, they require proper optimization on various devices. Another potential reason for the variance between rating and user review is that general users hardly focus on areas like perceived impacts, transparency, and technical functionality of the apps. Therefore, the user comments on the app store are not reliable in the case of determining the overall software quality characteristics of the apps.

\section{Findings and discussion}\label{findings}
\subsection{Limitations of reviewed apps}

\subsubsection{Feasibility of reviewed apps as CSA prevention programs}

Although there are plenty of apps that claim to provide CSA prevention information, most of the apps lack in fulfilling the key features (such as using game-based learning or serious games to teach children, involve parents in the education process, providing age and gender-specific education, etc.) that were deemed necessary by our study. Furthermore, delivering appropriate content to the target age group, creating content that will interest and encourage children and caregivers to learn, and evaluating the children's learning rate are some of the characteristics that are indispensable for an app to work effectively as a CSA prevention program~\cite{Kenny2012}. While apps may be targeted at different users, such as younger children, adolescents, or adults, their main goal is spreading awareness against the horrors of CSA. For an app to achieve the goal of being successful as a CSA prevention program, it must have both the software quality characteristics and the specific features required for spreading knowledge about CSA. Therefore, apps that cannot fulfill the measurement criteria determined by this study for rendering proper and precise CSA education are deemed unsuitable for being used as CSA prevention programs.

By analyzing the existing literature~\cite{Jones2020, Moon2017, Othman2015}, the CSA education-specific functionality sub-scale was devised that inquires about the most important features of a CSA education app such as parents involvement in education, using scenarios that children can relate to, including education about grooming, etc. Studies show that parent's involvement in the CSA education programs is crucial~\cite{Rudolph2018}. Also, parents themselves need to know all the crucial information about CSA and its prevention methods. Without the proper knowledge, they won't be able to protect their children or deal with post-abuse situations. Moreover, young children require parent's help to better understand what they are learning.
Among the apps we have reviewed, only one app (``The Cece Yara App") had a maximum of 7 out of 11 features. Six apps had 5 and more features and seven apps had less than 5 features. Among them, one app has fulfilled only one feature. Also, there was an app that did not satisfy any CSA education-specific features that were required. The app that has the maximum number of features, did not score very well in our overall rating. The reason for this inconsistency is that it has received a low score in other sub-scales of the rating scale like aesthetics, general app features, transparency, and subjective quality. A similar variance between software quality characteristics and CSA education-specific features was seen among the apps named ``Stop the Groomer", ``Helpio" and ``isafe". They also had more than average features but failed to meet the quality standard and thus, scored low in overall ratings. Another reason for this is the lack of more features and the features available being poorly optimized. On the flip side, the app ``Orbit Rescue" scored the highest rating in our modified rating scale. The reason for this is that this app provides all the necessary CSA education features along with required software quality characteristics. Another app ``Child Abuse Prevention" scored second-highest because it also possesses both the software qualities and the necessary features.


From the apps we reviewed, none fulfilled all of the crucial points necessary for rendering effective CSA education. For example, children with either physical or mental disabilities are more dependent on adults for their care. Hence, they stand a higher risk of being abused~\cite{Sanderson2004}, but only two (``The Cece Yara App" and ``Orbit Rescue") of the fourteen apps have mentioned disabled children. Even these two provided no information as to how disabled children can protect themselves or how their caregivers can keep them out of harm's way. One app included a character with a wheelchair as the main player (``Orbit Rescue"). And the other app (``The Cece Yara App" ) only informed adults about the increased risk of abuse for disabled children, but no information as to how they can be protected.
CSA education apps must be equally appealing and relatable to all of the target audience including children of all gender identities~\cite{Scholes2012}. However, the existing apps mainly focus on scenarios where female children are abused. In our study, we found that only three apps (``iSafe", ``Child Abuse Prevention" and ``Feel Safe") feature gender-specific abuse scenarios. This absence of knowledge regarding the situations where male children face the risk of abuse leaves them confused about which situations are dangerous for them and how to overcome those situations. 

Another important feature that the present apps fail to provide is proper age-specific CSA education. Small children have less attention span than older children. Also, younger children need special care and often require parental supervision for understanding what they are learning~\cite{He2001}. Therefore, apps targeted towards them need to have easily understandable content and a simple user interface. However, the simplistic content in the apps which are created for younger children won't be able to engage older children. Hence, the inclusion of age-specific content is indispensable. Only the app ``Feel Safe" provided age-specific scenarios of CSA and taught children how to prevent or deal with those situations. To protect themselves children need to recognize the various tactics that abusers apply to manipulate and exploit children~\cite{Scholes2012, Smallbone2001, Wurtele2014}. Unfortunately, only five of the fourteen apps discussed how to be safe from grooming tactics which are crucial for a child's safety. And only two of them, ``Stop the groomer" and ``Orbit Rescue" include proper preventative knowledge for children. Also, the app ``Child Abuse Info" provides such information for child care professionals.

Transparency is an extremely important part of any mobile app. Without proper credibility, no app can be considered trustworthy~\cite{Corral2014}. The questions we inquired about in our transparency sub-scale included user consent, the accuracy of store description, the legitimacy of the source, and the feasibility of achieving stated goals. Three out of 14 apps scored 3.5 and above out of 5 in the transparency sub-scale. These apps tried to fulfill the goals that they promised in their app store description. But only one app named ``Orbit Rescue" had published literature to support the fulfillment of their goal which is to assess the effectiveness of a game-based approach to teach CSA education to children aged 8–10 years~\cite{Scholes2014}. About 10 of 14 apps had related websites that provided more information about them and provided some credibility to them. Three apps (e.g., ``Bal Suraksha", ``Child Abuse Info", ``Game on POSCO") were part of government projects and thus had high legitimacy. Some apps were not explicit about the consent taken from the user. Moreover, some apps did not even have information about their privacy policy regarding user's personal information on their support websites or inside the app privacy policy page which raises a question about the authenticity of their intention of keeping user data private.

The most amazing part of having apps that teach about CSA is that children can learn by themselves from any location. However, for individual use of these apps as CSA prevention programs, they need to have all the necessary features and information. When child abuse prevention programs are integrated into school systems, their efficiency increases~\cite{Sanchez2019}. In a school, setting children are taught about CSA using small lessons across an extended period. These smaller lessons repeated for a longer period of time help children learn better. When children will get to revise what they learned at school using the apps at home, their understanding of the discussed matters will improve. Therefore, CSA education apps should be designed in a way that the apps can be used both at home and in a classroom setting. These apps should provide information about CSA and its prevention in an organized way that can be learned at home and also at school. Apps should also arrange their lectures in small lessons as children can develop key learnings progressively if lessons are provided in small portions and for a continued time period~\cite{Stieler-Hunt2014}. Only two apps we reviewed presented CSA education in a lesson-wise manner that can be used for both home and school-based education. This app is called ``Orbit Rescue". It was tested as part of the school curriculum in classroom settings and it evaluated the improvement of children's knowledge regarding CSA after using their app~\cite{Jones2020}.  Other apps that had prevention education divided into lessons were ``Child Abuse Prevention", ``Stop the Groomer" and ``Stewards of Children". The last one ``Stewards of Children" is intended for parents or caregivers. Among these three apps, the two apps ``Child Abuse Prevention" has detailed lesson-wise content that can be implemented in school. But, we found no evidence of these apps being tested at school.

Overall, it is clear that most of the apps present in app stores are not suitable for being used as CSA education programs. While a few may be able to teach children and parents individually, only one app could be deemed suitable for a school-based education program. These apps need to be improved both on their software qualities and CSA-specific features for being considered as potential CSA education programs.

\subsubsection{Potentiality of reviewed apps for inducing behavior changes about CSA education}

Without evaluating the impact of the apps on the target audience in real-life settings, it can not be determined whether these apps will induce behavior change among the users. After analyzing the apps we found that only one app has information regarding the impact of the app on children after using it. However, from our ratings acquired from the perceived impact sub-scale, it can be assumed that a few apps have the potential for bringing behavioral changes regarding the importance of prevention education for fighting CSA. Our modified rating scale has measured several factors for analyzing the feasibility of these apps in achieving the perceived impact we wanted. Behavior change towards CSA requires increasing awareness among users about the depth of the problem and spreading accurate knowledge concerning prevention~\cite{McKibbin2020}. These apps ideally should help to change both the attitude and intention of a user regarding CSA education. Furthermore, by using the apps users should be encouraged to seek help in cases of CSA. The apps (5/14) that did not meet these measurement criteria were deemed unsuitable for inducing awareness and help-seeking behavior regarding CSA education. However, our study showed that 8 out of 14 apps scored more than 4 in perceived impact on users sub-scale implying that these apps have the potential for inducing behavior changes of the users. Five of these apps are intended for children's use. Among these only one app (``Orbit Rescue") evaluated the impact of their app on children~\cite{Jones2020}. The researchers behind this app conducted a randomized control trial in school settings and evaluated children's knowledge before and after using the application. They found that children who completed the serious game for CSA education had increased knowledge and the serious game approach is effective for rendering CSA education. However, they did not analyze behavior change among the students. Hence, the actual behavior change impact cannot be determined. Such statistics suggest that while the commercial market has grown with an increasing number of CSA-related apps, ensuring their potentiality regarding inducing behavior change of the users requires properly administered study including children and adults.

\subsection{CSA app design considerations}
\subsubsection{Game-based approaches}
After analyzing 14 apps on CSA education, all of our ratings pointed to one conclusion: the most effective approach for teaching children sexual abuse prevention are game-based approaches such as gamification, game-based learning, and serious games. Gamification is an unconventional umbrella term that is used to express the use of game elements in a non-gaming practice to improve user experience and engagement~\cite{Pereira2014}. Game-based learning indicates the process and practice of learning using games~\cite{JanLPlass}. Whereas, serious games are custom-built games with a specific learning objective~\cite{Scholes2014}. Research indicates that systems built using gamification, game-based learning, or serious games are a more attractive, appealing, useful educational method for introducing prevention education to children~\cite{Desmet2015, Haruna2018, Scholes2014}. Through these kinds of systems, children can gain awareness about the need to protecting their bodies, learn how to recognize potential danger factors in their environment, and ask for help when in danger from trusted adults. More importantly, these methods help to form a link between home, social forces, and school~\cite{Shan2019}. A recent study shows that these game-based approaches help children to learn about prevention against this sensitive issue while making sure they do not get traumatized~\cite{Stieler-Hunt2014}.

The popularity of the apps which used games or game-based components is evident from the usage frequency of apps found by our raters. Eight apps had an average usage frequency of more than 10 times over 12 months. Five of these apps (named ``Orbit Rescue", ``Game on POSCO", ``Stop the Groomer", ``Child Abuse Prevention" and ``Feel Safe") were intended for children and all of them provided prevention education using either game-based components for teaching children or serious games focused on CSA prevention. Two of these apps (``Orbit Rescue", and ``Child Abuse Prevention") had a usage frequency of 10 to 50. One of these apps used the serious game method while the other had a game-based learning approach. We should mention that these two apps also scored the highest rating according to our scale. Moreover, the scores from the perceived impact sub-scale also suggest that apps that utilize the approach of teaching children via game-based components or games have more potential for inducing behavior change regarding CSA education. Five of the 8 apps which had more than 4 in this sub-scale rating were the same apps that used such approaches. The usage statistics and calculated ratings of such apps in this review are conclusive to the analysis of the user reviews from app stores as well, suggesting that the game-based learning, gamification, and serious games approaches are the most efficient and most preferred methods for educating children in CSA prevention.

\subsection{Design guidelines} %
This study shows that the existing CSA education apps available in the app stores are not fully suitable for rendering prevention education about CSA. They need further improvement on both software qualities (i.e., performance, aesthetics, usability) and CSA-specific features. Therefore, future development must consider the limitations found in the existing apps that have been pointed out in our study. Developers can test their apps using our rating scale to evaluate the apps' effectiveness. The CSA education rating scale devised in our study will also be helpful to developers for deciding the most important features necessary for a CSA education app.

Following the reviews and ratings of the existing CSA education apps and evaluating the recent research on effective methods for rendering prevention education, we believe that some additional points should be considered by developers before designing an app for this purpose. Children tend to learn better when they find the learning constituents relatable to their life~\cite{Hamari2016}. Therefore, the apps' contents must include scenarios from real life. Game-based apps are better for educating children about sensitive issues like sexual abuse~\cite{Schoech2013}. Hence, developers should focus on game-based learning, gamification or serious game approaches for creating future CSA education apps. Furthermore, user customization features should be increased so that children can easily get into the characters in the apps. The typical quiz-based learning apps often fail to engage children while teaching~\cite{Egenfeldt-nielsen2013, Kirriemuir2004}. Therefore, designers should consider ways where children can learn through playing and observing. If quiz methods are required for evaluation, children should be given the scope for fixing their mistakes in a way that goes beyond rote learning~\cite{Stieler-Hunt2014}. Research shows that children who possess healthy self-esteem are more likely to accept the message delivered in CSA prevention education programs~\cite{Sanderson2004}. So developers should focus on creating apps that will promote behavioral change and help build a positive self-concept in children. The inclusion of parents or caregivers in the same program as children is specifically essential for teaching younger children~\cite{Kang2020}. This also helps children to choose their trusted adults and build healthy communication with them. Thus the apps must include information for both the children and parents. Also, the apps must promote the necessary knowledge for the trusted adults, so that they can act appropriately when an abusive incident is disclosed. Because when the adults will understand the dangers of CSA and be properly educated in prevention education, they will be more eager to educate their children using the CSA  education apps.




\section{Conclusions and future work}\label{conclusion}
In this study, we have systematically reviewed the existing CSA education apps available in the app stores. For selecting apps that meet our study criteria we formulated specific keywords and inclusion criteria. From the initial 191 apps, we finally selected 14 apps for review. For analyzing the selected apps, we devised an app rating scale by modifying the existing rating scales keeping in mind the specific needs of evaluating the CSA  education apps. The verdict drawn from the use of our rating scale on the selected CSA apps and the analysis of the results showed that in its current state, the majority of apps in app stores do not meet satisfactory features required to teach children about CSA and its prevention. The flaws observed in this review show that not only important features are needed but also the software quality of these apps need to be improved. Results also indicate that children prefer apps that use game-based learning approaches for teaching CSA education. This study also provides the necessary knowledge to developers and individuals regarding software quality characteristics and CSA education. Furthermore, it points at the possible directions of the advancement of research and development in CSA education apps. Moreover, individuals can gain insight from this study about what features a CSA education app must have to be truly useful. With the help of this study, developers will be able to design apps that will be eligible for motivating individuals to properly address CSA and become more aware of the importance of learning about CSA education. 

The opinion from a child expert can help to improve the rating scale for CSA education apps and provide a comprehensive CSA  education structure for children. Therefore, future research will focus on including a specialist's perspective regarding this topic for a more credible research outcome. Another future recommendation for this study would be to evaluate these apps with both children and adults in real-life settings. The necessary feedback from respective users can enhance the possibility of including more features. Also, from such a study, the effectiveness of these apps for providing CSA  education would become clear. Our study included only those CSA apps that are available in English and Bangla languages since these are the languages that authors are familiar with. As a consequence, even though we reviewed an enormous number of CSA apps, we may have missed some apps because of our language barrier. Another factor that needs to be noted is that there may exist some apps that could not be accessed due to regional restrictions. Also, Since no prior research has been done on searching and reviewing the CSA  education apps, the searching methods used in our review were modeled after those previously conducted in other research area~\cite{Milne-Ives2020, Rivera2016}. Thus, another research direction would be expanding the study regardless of searching criteria, language variations, and regional restrictions of the app stores.

\section*{Acknowledgements}\label{Acknowledgement}
The authors would like to thank Sumaiya Nazmin Nishu for her help with the app ratings. 
\section*{Funding Source}\label{Funding Source}
This research did not receive any specific grant from funding agencies in the public, commercial, or non-profit sectors.


\pagebreak

\bibliographystyle{elsarticle-num.bst} 
\bibliography{sample-base}

\begin{thebibliography}{10}
\expandafter\ifx\csname url\endcsname\relax
  \def\url#1{\texttt{#1}}\fi
\expandafter\ifx\csname urlprefix\endcsname\relax\def\urlprefix{URL }\fi
\expandafter\ifx\csname href\endcsname\relax
  \def\href#1#2{#2} \def\path#1{#1}\fi

\bibitem{Singh2014}
M.~Singh, S.~Parsekar, S.~Nair, {An epidemiological overview of child sexual
  abuse}, Journal of Family Medicine and Primary Care 3~(4) (2014) 430--435.

\bibitem{Csorba2004}
R.~Csorba, R.~P{\'{o}}ka, P.~Sz{\'{e}}kely, A.~Borsos, L.~Balla,
  {\'{E}}.~Ol{\'{a}}h, {Child sexual abuse}, Orvosi hetilap 145~(5) (2004)
  223--227.

\bibitem{Goldman2000}
J.~D. Goldman, U.~K. Padayachi, {Some methodological problems in estimating
  incidence and prevalence in child sexual abuse research}, The Journal of Sex
  Research 37~(4) (2000) 305--314.

\bibitem{WorldHealthOrganization2020}
{World Health Organization}, {Global status report on preventing violence
  against children}, Geneva: World Health Organization, 2020.

\bibitem{Darknesstolight}
{Darkness to light}, Darkness to light,
  \url{https://www.d2l.org/the-issue/statistics/} (accessed 13 January 2021).

\bibitem{Cho2015}
Y.-R. Cho, J.-E. Kim, K.-M. Park, {Differences in the Characteristics of Sexual
  Abuse Victimization between Low- and High-Grade Elementary School Children
  and Correlations among the Characteristics}, Journal of Korean Academy of
  Community Health Nursing 26~(2) (2015) 119--127.

\bibitem{Sanderson2004}
J.~Sanderson, {Child-focused sexual abuse prevention programs How effective are
  they in preventing child abuse?}, Crime and Misconduct Commission Brisbane,
  Qld, 2004.

\bibitem{Daray2016}
F.~M. Daray, S.~M. Rojas, A.~J. Bridges, C.~L. Badour, L.~Grendas, D.~Rodante,
  S.~Puppo, F.~Rebok, {The independent effects of child sexual abuse and
  impulsivity on lifetime suicide attempts among female patients}, Child Abuse
  \& Neglect 58 (2016) 91--98.

\bibitem{Cullen2020}
O.~Cullen, K.~{Zug Ernst}, N.~Dawes, W.~Binford, G.~Dimitropoulos, {“Our Laws
  Have Not Caught up with the Technology”: Understanding Challenges and
  Facilitators in Investigating and Prosecuting Child Sexual Abuse Materials in
  the United States}, Laws 9~(4) (2020) 1--19.

\bibitem{He2001}
M.~Hébert, F.~Lavoie, C.~Piché, M.~Poitras, {Proximate effects of a child
  sexual abuse prevention program in elementary school children}, Child Abuse
  \& Neglect 25 (2001) 505--522.

\bibitem{Holloway2018}
J.~L. Holloway, M.~L. Pulido, {Sexual Abuse Prevention Concept Knowledge: Low
  Income Children Are Learning but Still Lagging}, Journal of Child Sexual
  Abuse 27~(6) (2018) 642--662.

\bibitem{Rudolph2018}
J.~Rudolph, M.~J. Zimmer-Gembeck, D.~C. Shanley, R.~Hawkins, {Child Sexual
  Abuse Prevention Opportunities: Parenting, Programs, and the Reduction of
  Risk}, Child Maltreatment 23~(1) (2018) 96--106.

\bibitem{BalSurakshaApp}
{Mobile Seva}, {Bal Suraksha},
  \url{https://play.google.com/store/apps/details?id=com.cdac.safe_spaces_children_English},
  (accessed 28 January 2021).

\bibitem{ElementsApp}
{Academy for Professional Excellence}, {Elements of Child Sexual Abuse},
  \url{https://play.google.com/store/apps/details?id=com.andromo.dev88207.app127483},
  (accessed 21 December 2020).

\bibitem{Stieler-Hunt2014}
C.~Stieler-Hunt, C.~M. Jones, B.~Rolfe, K.~Pozzebon, {Examining key design
  decisions involved in developing a serious game for child sexual abuse
  prevention}, Frontiers in psychology 5 (2014) 1--10.

\bibitem{Noei2017}
E.~Noei, M.~D. Syer, Y.~Zou, A.~E. Hassan, I.~Keivanloo, {A study of the
  relation of mobile device attributes with the user-perceived quality of
  Android apps}, Empirical Software Engineering 22~(6) (2017) 3088--3116.

\bibitem{Rivera2016}
J.~Rivera, A.~McPherson, J.~Hamilton, C.~Birken, M.~Coons, S.~Iyer, A.~Agarwal,
  C.~Lalloo, J.~Stinson, {Mobile Apps for Weight Management: A Scoping Review},
  JMIR mHealth and uHealth 4~(3) (2016) e87.

\bibitem{Tricco2018}
A.~Tricco, E.~Lillie, W.~Zarin, K.~O’Brien, H.~Colquhoun, D.~Levac, D.~Moher,
  M.~Peters, T.~Horsley, L.~Weeks, S.~Hempel, E.~Akl, C.~Chang, J.~Mcgowan,
  L.~Stewart, L.~Hartling, A.~Aldcroft, M.~Wilson, C.~Garritty, S.~Straus,
  Prisma extension for scoping reviews (prisma-scr): Checklist and explanation,
  Annals of Internal Medicine 169 (09 2018).

\bibitem{Stawarz2015}
K.~Stawarz, A.~L. Cox, A.~Blandford, {Beyond self-tracking and reminders:
  Designing smartphone apps that support habit formation}, in: Conference on
  Human Factors in Computing Systems - Proceedings, 2015, pp. 2653--2662.

\bibitem{Faux2016}
F.~Faux, R.~Bastide, N.~Souf, R.~Zgheib, {Smartphone-Based Collaborative System
  for Wounds Tracking Open Archive TOULOUSE Archive Ouverte (OATAO)}, in:
  eTELEMED 2016, 2016, pp. 104--109.

\bibitem{Poon2015}
T.~W.~K. Poon, M.~R. Friesen, {Algorithms for size and color detection of
  smartphone images of chronic wounds for healthcare applications}, IEEE Access
  3 (2015) 1799--1808.

\bibitem{Friesen2013}
M.~R. Friesen, C.~Hamel, R.~D. McLeod, {A mHealth application for chronic wound
  care: Findings of a user trial}, International journal of environmental
  research and public health 10~(11) (2013) 6199--6214.

\bibitem{Vos-Draper2013}
T.~Vos-Draper, {Poster 29 Wireless Seat Interface Pressure Mapping on a
  Smartphone: Feasibility Study in Users with SCI}, Archives of Physical
  Medicine and Rehabilitation 94~(10) (2013) e21--e22.

\bibitem{Kabir2021}
M.~A. Kabir, S.~S. Rahman, M.~M. Islam, S.~Ahmed, C.~Laird, Mobile apps for
  foot measurement in pedorthic practice: Scoping review, JMIR Mhealth Uhealth
  9~(3) (2021) e24202.

\bibitem{Stoyanov2015}
S.~R. Stoyanov, L.~Hides, D.~J. Kavanagh, O.~Zelenko, D.~Tjondronegoro,
  M.~Mani, {Mobile App Rating Scale: A New Tool for Assessing the Quality of
  Health Mobile Apps}, JMIR mHealth uHealth 3~(1) (2015) e27.

\bibitem{Stoyanov2016}
S.~R. Stoyanov, L.~Hides, D.~J. Kavanagh, H.~Wilson, {Development and
  validation of the user version of the mobile application rating scale
  (uMARS)}, JMIR mHealth uHealth 4~(2) (2016) 1--5.

\bibitem{Moon2017}
K.~J. Moon, K.~M. Park, Y.~Sung, {Sexual Abuse Prevention Mobile Application
  (SAP{\_}MobAPP) for Primary School Children in Korea}, Journal of Child
  Sexual Abuse 26~(5) (2017) 573--589.

\bibitem{Speckyboy2020}
Speckyboy, {The Aesthetics of a Successful Mobile App: Lessons From Three Top
  Grossing Apps}, \url{https://speckyboy.com/the-aesthetics-of-a-
  successful-mobile-app-lessons-from-three-top-grossing-apps/}, (accessed 25
  December 2020).

\bibitem{Hamari2016}
J.~Hamari, D.~J. Shernoff, E.~Rowe, B.~Coller, J.~Asbell-Clarke, T.~Edwards,
  {Challenging games help students learn: An empirical study on engagement,
  flow and immersion in game-based learning}, Computers in Human Behavior 54
  (2016) 170--179.

\bibitem{Torous2018}
J.~Torous, J.~Nicholas, M.~E. Larsen, J.~Firth, H.~Christensen, {Clinical
  review of user engagement with mental health smartphone apps: Evidence,
  theory and improvements}, Evidence Based Mental Health 21~(3) (2018)
  116--119.

\bibitem{Kekalainen2005}
A.~Kek{\"{a}}l{\"{a}}inen, A.~Kaikkonen, A.~Kankainen, M.~Cankar, T.~Kallio,
  {Usability Testing of Mobile Applications: A Comparison between Laboratory
  and Field Testing}, Journal of Usability Studies 1~(1) (2005) 4--16.

\bibitem{Walsh2012}
K.~Walsh, L.~Brandon, {Their Children's First Educators: Parents' Views About
  Child Sexual Abuse Prevention Education}, Journal of Child and Family Studies
  21~(5) (2012) 734--746.

\bibitem{Kang2020}
S.~R. Kang, S.~J. Kim, K.~A. Kang, {Effects of Child Sexual Abuse Prevention
  Education Program Using Hybrid Application (CSAPE-H) on Fifth-Grade Students
  in South Korea}, Journal of School Nursing (2020) 1--12.

\bibitem{Bennett2020}
N.~Bennett, W.~O. Donohue, {Identifying Grooming of Children for Sexual Abuse :
  Gender Effects and Increased False Positives from Internet Information},
  International Journal of Psychology and Psychological Therapy (2020)
  133--145.

\bibitem{Scholes2012}
L.~Scholes, C.~Jones, C.~Stieler-hunt, B.~Rolfe, K.~Pozzebon, {The Teachers '
  Role in Child Sexual Abuse Prevention Programs : Implications for Teacher
  Education}, Australian Journal of Teacher Education 37~(11) (2012).

\bibitem{chang2020}
K.~C. Chang, R.~N. Zaeem, K.~S. Barber, Is your phone you? how privacy policies
  of mobile apps allow the use of your personally identifiable information, in:
  2020 Second IEEE International Conference on Trust, Privacy and Security in
  Intelligent Systems and Applications (TPS-ISA), IEEE Computer Society, 2020,
  pp. 256--262.

\bibitem{Tait2018}
R.~J. Tait, J.~J.~L. Kirkman, M.~P. Schaub, {A Participatory Health Promotion
  Mobile App Addressing Alcohol Use Problems (The Daybreak Program ): Protocol
  for a Randomized Controlled Trial}, JMIR research protocols 7~(June) (2018).

\bibitem{Jones2020}
C.~Jones, L.~Scholes, B.~Rolfe, C.~Stieler-Hunt, {A serious-game for child
  sexual abuse prevention: An evaluation of orbit}, Child Abuse and Neglect 107
  (2020).

\bibitem{Milne-Ives2020}
M.~Milne-Ives, C.~LamMEng, C.~de~Cock, M.~H. van Velthoven, E.~M. Ma, {Mobile
  apps for health behavior change in physical activity, diet, drug and alcohol
  use, and mental health: Systematic review}, JMIR mHealth uHealth 8~(3)
  (2020).

\bibitem{Cronbach1951}
L.~J. Cronbach, {Coefficient alpha and the internal structure of tests},
  Psychometrika 16~(3) (1951) 297--334.

\bibitem{Gliem2003}
J.~a. Gliem, R.~R. Gliem, {Calculating, Interpreting, and Reporting Cronbach's
  Alpha Reliability Coefficient for Likert-Type Scales}, 2003 Midwest Research
  to Practice Conference in Adult, Continuing, and Community Education (2003)
  82--88.

\bibitem{Sawa2007}
J.~Sawa, T.~Morikawa, {Inter-rater reliability for multiple raters in clinical
  trials of ordinal scale}, Drug Information Journal 41~(5) (2007) 595--605.

\bibitem{Hallgren2012}
K.~A. Hallgren, {Computing Inter-Rater Reliability for Observational Data: An
  Overview and Tutorial}, Tutorials in Quantitative Methods for Psychology
  8~(1) (2012) 23--34.

\bibitem{Koo2016}
T.~K. Koo, M.~Y. Li, {A Guideline of Selecting and Reporting Intraclass
  Correlation Coefficients for Reliability Research}, Journal of Chiropractic
  Medicine 15~(2) (2016) 155--163.

\bibitem{BEHNAMNIA2020}
N.~Behnamnia, A.~Kamsin, M.~A.~B. Ismail, A.~Hayati, The effective components
  of creativity in digital game-based learning among young children: A case
  study, Children and Youth Services Review 116 (2020) 105227.

\bibitem{Guzman2018}
E.~{Guzman}, L.~{Oliveira}, Y.~{Steiner}, L.~C. {Wagner}, M.~{Glinz}, User
  feedback in the app store: A cross-cultural study, in: 2018 IEEE/ACM 40th
  International Conference on Software Engineering: Software Engineering in
  Society (ICSE-SEIS), 2018, pp. 13--22.

\bibitem{Vasa2012}
R.~Vasa, L.~Hoon, K.~Mouzakis, A.~Noguchi, {A preliminary analysis of mobile
  app user reviews}, in: Proceedings of the 24th Australian Computer-Human
  Interaction Conference, OzCHI 2012, OzCHI '12, Association for Computing
  Machinery, NY, USA, 2012, pp. 241--244.

\bibitem{Kenny2012}
M.~C. Kenny, S.~K. Wurtele, {Preventing Childhood Sexual Abuse : An Ecological
  Approach}, Journal of Child Sexual Abuse 21~(October 2013) (2012) 361--367.

\bibitem{Othman2015}
A.~Othman, W.~A.~J. {Wan Yahaya}, B.~Muniandy, {Integration of persuasive
  multimedia in designing learning application for child sexual abuse}, Jurnal
  Teknologi 77~(29) (2015) 97--101.

\bibitem{Smallbone2001}
S.~Smallbone, R.~Wortley, {Child sexual abuse in Queensland : Offender
  characteristics and modus operandi}, Trends and Issues in Crime and Criminal
  Justice 193~(January) (2001).

\bibitem{Wurtele2014}
S.~K. Wurtele, {Preventing Sexual Abuse of Children in the Twenty-First Century
  : Preparing for Challenges and Opportunities}, Journal of Child Sexual Abuse
  18~(May) (2014) 1--18.

\bibitem{Corral2014}
L.~Corral, A.~Sillitti, G.~Succi, {Defining Relevant Software Quality
  Characteristics from Publishing Policies of Mobile App Stores}, in: M.~Awan,
  I.and~Younas, X.~Franch, C.~e. Quer (Eds.), {Mobile Web Information Systems},
  Vol. 8640, Springer, Cham, 2014, pp. 205--217.

\bibitem{Scholes2014}
L.~Scholes, C.~Jones, C.~Stieler-Hunt, B.~Rolfe, {Serious games for learning:
  Games-based child sexual abuse prevention in schools}, International Journal
  of Inclusive Education 18~(9) (2014) 934--956.

\bibitem{Sanchez2019}
A.~Sánchez, M.~Favero, Effectiveness of programs for the prevention of child
  sexual abuse: A comprehensive review of evaluation studies, European
  Psychologist 25 (2019) 1--15.

\bibitem{McKibbin2020}
G.~McKibbin, C.~Humphreys, {Future directions in child sexual abuse prevention:
  An Australian perspective}, Child Abuse and Neglect 105~(December 2019)
  (2020) 104422.

\bibitem{Pereira2014}
P.~Pereira, E.~Duarte, F.~Rebelo, P.~Noriega, A review of gamification for
  health-related contexts, in: A.~Marcus (Ed.), Design, User Experience, and
  Usability. User Experience Design for Diverse Interaction Platforms and
  Environments, Vol. 8518 of Lecture Notes in Computer Science, Springer, Cham,
  2014, pp. 742--753.

\bibitem{JanLPlass}
J.~L. Plass, B.~D. Homer, C.~K. Kinzer, Foundations of game-based learning,
  Educational Psychologist 50~(4) (2015) 258--283.

\bibitem{Desmet2015}
A.~Desmet, R.~Shegog, D.~{Van Ryckeghem}, G.~Crombez, I.~{De Bourdeaudhuij}, {A
  Systematic Review and Meta-analysis of Interventions for Sexual Health
  Promotion Involving Serious Digital Games}, Games Health Journal 4~(2) (2015)
  78--90.

\bibitem{Haruna2018}
H.~Haruna, X.~Hu, S.~K.~W. Chu, R.~R. Mellecker, G.~Gabriel, P.~S. Ndekao,
  {Improving sexual health education programs for adolescent students through
  game-based learning and gamification}, International Journal of Environmental
  Research and Public Health 15~(9) (2018).

\bibitem{Shan2019}
Y.~Shan, {Protect Me: An interactive mobile educational tool for children to
  help and prevent sexual abuse}, Master's thesis, Rochester Institute of
  Technology (2019).

\bibitem{Schoech2013}
D.~Schoech, J.~F. Boyas, B.~M. Black, N.~Elias, { Gamification for Behavior
  Change : Lessons from Developing a Social , Multiuser , Web-Tablet Based
  Prevention Game for Youths}, Journal of Technology in Human Services
  31~(August 2013) (2013) 197--217.

\bibitem{Egenfeldt-nielsen2013}
S.~Egenfeldt-nielsen, {Beyond Edutainment: Exploring the Educational Potential
  of Computer Games}, Lulu.com, Købmagergade 11A, 4. floor 1150, Copenhagen,
  2013.

\bibitem{Kirriemuir2004}
J.~Kirriemuir, A.~McFarlane, {Literature Review in Games and Learning}, Nesta
  Futurelab Bristol, 2004.

\end{thebibliography}





\end{document}